# Unified image formation theory for microscopy and optical coherence tomography in 4-D space-time

Naoki Fukutake[1,2]*, Shuichi Makita[2], Yoshiaki Yasuno[2]

[1]Nikon Corporation; 471 Nagaodai-cho, Sakae-ku, Yokohama City, Kanagawa, 244-8533, Japan

[2]University of Tsukuba; Tennodai 1-1-1, Tsukuba, Ibaraki, 305-8573 Japan

*Correspondence to: Naoki.Fukutake@nikon.com

## Abstract

We construct an image formation theory that covers the majority of optical microscopy techniques that use diverse coherent or incoherent light-matter interactions. The theories of individual microscopy methods could not previously be connected with other systems because of the absence of a common theoretical framework. Using the general principles of quantum physics and applying a four-dimensional representation, we unify the image formation theories of optical systems ranging from classical microscopy to cutting-edge instruments into a single framework in which light is replaced with quantum fields and the interactions are represented using double-sided Feynman diagrams. Our universal methodology requires a four-dimensional aperture that enables sufficient understanding of the associations between the different imaging types and interprets image formation appropriately for all systems, including optical coherence tomography.

## Introduction

The essentials of image formation and the origins of optical resolution in imaging systems were established before the emergence of modern physics, which is based on relativity and quantum theory. This classical image formation theory provides a well-known expression for the optical resolution of $\lambda/2\mathrm{NA}$, where λ and NA are the wavelength and the numerical aperture, respectively. The classical theory addresses image formation effectively for classical optical microscopy techniques, including bright-field (*1*,*2*), dark-field (*3*), phase-contrast (*4*,*5*), and differential interference microscopy (*6*). However, in terms of the light-matter interactions used to form an image, only the lowest order interactions ($\chi^{(1)}$ interactions) such as linear absorption, where $\chi^{(i)}$ is an *i*-th order susceptibility, follow classical theory (*7*). In higher order interactions, including fluorescence (FL, which is a $\chi^{(3)}$ interaction), microscopy does not necessarily follow the optical resolution formula $\lambda/2\mathrm{NA}$ (*8*, *9*). In this case, for FL microscopy alone, a phenomenological formula has been used as a solution in which light is treated not as an amplitude but as an intensity (*10*). For other microscopy methods with higher order interactions, almost no precision image formation theory exists, and the optical resolution has simply been estimated for each technique individually. After the emergence of lasers in particular, a diverse range of interactions, including two-photon excited FL (TPEF), second-/third-harmonic generation (SHG/THG), and coherent Raman scattering (CRS), have been applied to optical microscopy (*11*-*16*). Despite the rapid growth in photonics technology, classical theory and these individual approaches are still used to express optical resolution.

Each individual approach is useful in practice for interpretation of experimental results, including estimation of the optical resolution in specific microscopy methods. However, these individual approaches do not provide relationships between different microscopy types and cannot describe the essentials of their light-matter interactions. In particular, the coherent and incoherent interactions used in microscopy are described as entirely different optical processes. Because FL is also an electromagnetic wave with a phase, its treatment as an intensity is incompatible with some optical phenomena, including interference. Moreover, most individual approaches, including that for optical coherence tomography (OCT), do not have exact image formation equations (*17*). Furthermore, if a novel imaging technique emerges, these individual approaches cannot be applied to the new imaging system.

It is natural to consider devising a unified image formation expression suitable for all imaging systems. To evaluate and compare the optical resolutions of each microscopy method, a common theoretical framework has been formulated that covers both classical and modern microscopy (*7*). In Ref. (*7*), the image formation methods for almost all far-field microscopies are addressed in a universal manner using three-dimensional quantum optics. Although this common framework provides insight into the origins of the optical resolution, it does not work for some imaging systems, including OCT and pump-probe (PP) microscopy with a delay line (*17, 18*). Because the interactions that cannot be explained using modern physics never occur during microscopy, a systematic approach must be formulated, even if the different imaging system types, including FL, TPEF, SHG, THG, CRS, and PP microscopies, and OCT, appear entirely distinct and unrelated. Recently, as more cutting-edge imaging system types have been developed, it has become desirable to update image formation theory to produce a modern theory that provides a new framework to enable a unified interpretation of the essentials of image formation.

One major issue is determination of how to formulate this modern image formation theory. The unified framework must be constructed by completing the following steps. First, we expand a coordinate system into four dimensions in the same manner as relativity because some imaging system types, e.g., PP microscopy and OCT, will inevitably require an additional dimension to express their phenomena. Second, we generalize the lowest order interaction (diffraction) with respect to an arbitrary order interaction (phase matching (PM)), i.e., diffraction is redefined as the lowest order of PM. Because all the interactions can be described using double-sided Feynman diagrams (*19-21*), the diagrammatical methodology with linear/nonlinear susceptibility $\chi^{(i)}$ can be incorporated into the formula. Third, we quantize the light field to introduce the vacuum field, because the vacuum field acts as an excitation field in incoherent interactions such as FL. These incoherent interactions will be treated as PM with the vacuum field, which clarifies that the nature of the incoherence in FL stems from the vacuum random phase. By treating both coherent and incoherent interactions in a consistent manner, it is found that the vacuum field helps to improve the optical resolution (*7*). Finally, we consider a local oscillator (LO), i.e., intense light that is not influenced by sample structures. The presence or absence of this LO affects the image formation characteristics, which then enables systematic treatments to be applied for all imaging systems.

In this paper, we demonstrate how a unified image formation theory is constructed by applying modern physics to imaging systems. First, we introduce a universal image formation formula, from which image formation formulas can be derived for all imaging systems. Next, we define a four-dimensional (4D) aperture that is derived from this universal formula and a new 4D pupil function concept. The 4D aperture determines the spatial-temporal resolution of each imaging system and enables comparison of the optical resolutions of all imaging systems. Finally, we confirm that our framework includes each individual existing approach.

## Results

### Theoretical framework

To construct a universal image formation formula that can represent all optical imaging systems, we first formulate the required framework using modern physics: specifically, we use quantum theory in 4D space-time.

#### *1. Four-dimensional Fourier optics*

We construct a 4D Fourier optics system based on conventional three-dimensional (3D) Fourier optics (*22*) for imaging systems. By adding an angular frequency of light to a wavevector, we can consider a 4D wavevector $(k_x, k_y, k_z, \omega/c)$, where $\boldsymbol{k} = (k_x, k_y, k_z)$ represents the 3D wavevector, ω is the angular frequency, $c$ is the speed of light, and the dispersion relation $|\boldsymbol{k}| = \omega/c$ is satisfied. As part of a conic surface $\delta\{|\boldsymbol{k}|^2 - (\omega/c)^2\} = (c/2\omega)[\delta\{|\boldsymbol{k}| - (\omega/c)\} + \delta\{|\boldsymbol{k}| + (\omega/c)\}]$ in the 4D wavevector domain (*23*), a 4D pupil function $P(\boldsymbol{k}, \omega/c)$ of an optical system with an NA and a bandwidth can be defined as shown in Fig. 1A. The 4D pupil function forms a Fourier transform pair with a 4D amplitude spread function (ASF) $h(x, y, z, ct)$ in the 4D space-time domain: $h(\boldsymbol{x}, ct) = \int P(\boldsymbol{k}, \omega/c) e^{i(\boldsymbol{k}\cdot\boldsymbol{x} - \omega t)} d^3\boldsymbol{k} d\omega$, where $\boldsymbol{x} = (x, y, z)$ and $t$ represent a 3D space and the time in the 4D object coordinates, respectively. Figure 1B illustrates an example of the 4D ASF, which can be expressed in the form of the retarded Green's function $G_{\mathrm{ret}}\{(\boldsymbol{x}, ct) \leftarrow (\boldsymbol{0}, 0)\} = h(\boldsymbol{x}, ct)$, where $\boldsymbol{0} = (0, 0, 0)$. A detailed description of the Green's function (*24*) is given in Method. Hereafter, we use the relation $(\boldsymbol{k}, \omega/c) = 2\pi(\boldsymbol{f}, \nu/c)$ on a case-by-case basis, where $\boldsymbol{f} = (f_x, f_y, f_z)$ is a 3D spatial frequency, $\nu$ is the light frequency, and $(\boldsymbol{f}, \nu/c)$ is referred to as a 4D frequency.

#### 2. *Phase matching in the interaction represented by a Feynman diagram*

Bragg diffraction in 3D $\chi^{(1)}$-gratings can be generalized using a superordinate concept, i.e., the PM in $\chi^{(i)}$-gratings (*21, 25*). For any order of $\chi^{(i)}$-derived interactions, the PM becomes a fundamental phenomenon related to the optical resolution. Figure 2A shows an example of the PM for sum-frequency generation (SFG), in which two excitation beams are incident on a $\chi^{(2)}$-grating with grating vector $G$ that represents one of the Fourier components of the $\chi^{(2)}$ distribution, generating an SFG signal wave with wavevector $k_{\mathrm{sig}}$. Note that the two plane waves with wavevectors $k_{\mathrm{ex1}}$ and $k_{\mathrm{ex2}}$ represent the Fourier components of the two excitation beams when focused by a lens. The signal wave only appears if the PM condition $k_{\mathrm{ex1}} + k_{\mathrm{ex2}} + G = k_{\mathrm{sig}}$ is satisfied.

All interaction types can be expressed using double-sided Feynman diagrams. Typical example diagrams are shown in Fig. 2B. These diagrams describe the time evolution of a density matrix for molecules that interact with excitation fields, where the time evolves from the bottom toward the top. In the diagrams, solid arrows represent excitation fields from light sources, dotted arrows represent those for the vacuum fields, and we do not draw an arrow (a wavy arrow going outward from the top on the left side) for the signal field. The left and right sides of each diagram correspond to the positive and negative frequencies of the excitation fields, respectively. The arrows going inward and outward represent excitation and de-excitation, respectively. In incoherent interactions, the vacuum field acts as one of the excitation fields, which causes the PM to include the vacuum field. The dotted arrow that describes the vacuum always goes outward, i.e., it deexcites. In $\chi^{(i)}$-derived interactions, where $i$ represents an interaction order, $i$ excitation fields are engaged in the PM. After the interaction, the molecules that are distributed as a $\chi^{(i)}$-grating radiate signal fields that only interfere constructively in the direction that meets the PM condition.

## 3. *Field quantization for incorporation of the vacuum*

To treat coherent and incoherent interactions within the same framework, we quantize the fields by replacing the classical fields with creation/annihilation operators, which naturally include the vacuum field concept (*19, 26*). One typical incoherent interaction is FL, which is also illustrated schematically in Fig. 2B. In FL, because two excitation fields and one vacuum field are responsible for signal generation, we consider the PM of the $\chi^{(3)}$-derived interaction that includes the vacuum field. Specifically, the incoherent nature of FL is considered to originate from a random phase of the vacuum field, i.e., the zero-photon state.

In field quantization, operators are usually defined in the 3D wavevector domain, e.g., $\hat{a}(k_x, k_y, k_z)$, as depicted in the diagram in Fig. 2B. Furthermore, to use the diagram for image formation theory, we transform the operators for the 3D wavevector domain into those for the 4D space-time domain (*27*), i.e., $\hat{a}(x, y, z, ct)$, by using the 4D pupil function, as exemplified for the stimulated Raman gain (SRG) case in Fig. 2C (see Method for further details). Expressing the diagram in the space-time domain is an essential technique in our theory that provides a diagram rule for calculation of the $i$-th order polarization operator $\hat{P}_{\mathrm{o}}^{(i)}(\boldsymbol{x}, t)$. $\hat{P}_{\mathrm{o}}^{(i)}(\boldsymbol{x}, t)$ is the product of a molecular density distribution $N(\boldsymbol{x})$ and the polarization of a molecule at $\boldsymbol{x}$, denoted by $\hat{p}^{(i)}(\boldsymbol{x}, t)$, where $\hat{p}^{(i)}(\boldsymbol{x}, t)$ is calculated by following the diagram rule described in Method (Single-molecule polarization). Figure 2D shows an example of the time evolution of $\hat{p}^{(i)}(\boldsymbol{x}, t)$ in SRG, i.e., the $\chi^{(3)}$ interaction, which corresponds to the diagram in Fig. 2C. Here, $\hat{p}^{(i)}(\boldsymbol{x}, t)$ is excited by the three pulses at $t_1$, $t_2$, and $t_3$, where $t$ represents a signal field emission time, and the signal field is emitted by $\hat{p}^{(i)}(\boldsymbol{x}, t)$. Using $N(\boldsymbol{x})$, the susceptibility distribution $\chi^{(i)}(\boldsymbol{x})$ can be decomposed into $N(\boldsymbol{x})$ and the space-invariant component $\chi_{\mathrm{p}}^{(i)}$ because $\chi^{(i)}(\boldsymbol{x}) = N(\boldsymbol{x})\chi_{\mathrm{p}}^{(i)}$.

## 4. *Model of optical imaging system with/without LO*

We define microscopy and tomography as systems that involve 1) an excitation system, 2) light-matter interaction in the sample, and 3) a signal collection (detection) system (*1-6, 8, 9, 11-16, 18, 28, 29*). Some imaging systems also include 4) a reference arm (RA) (*17, 30, 31, 32*). Figure 3 shows examples of imaging systems. Our model covers all forms of far-field optical imaging systems that do not use *a priori* information about the sample structure, including laser/sample-stage scanning microscopy, Kohler illumination microscopy with a two-dimensional detector, and OCT.

In our model, the object is $N(\boldsymbol{x})$, i.e., the density distribution of the molecules in the sample that emits the signal to be observed. After the excitation light (which includes the vacuum) interacts with the molecules distributed as $N(\boldsymbol{x})$, the signal light emerges and then propagates toward a detector placed at $z_{\mathrm{d}} = 0$ in a detector coordinate $\boldsymbol{x}_{\mathrm{d}} = (x_{\mathrm{d}}, y_{\mathrm{d}}, z_{\mathrm{d}})$, where the image plane $z_{\mathrm{d}} = 0$ is conjugate to the object plane $z = 0$. Note that the light-matter interaction can be decomposed into numerous PMs in the frequency domain. Here, the object and detector coordinates are represented by $\boldsymbol{x}$ and $\boldsymbol{x}_{\mathrm{d}}$, respectively, whereas the coordinate in the image is expressed as $\boldsymbol{x}' = (x', y', z')$ in the later sections.

Next, we elaborate on the LO. The LO is present in some imaging system types and is defined as a field that is not influenced by the molecular distribution $N(\boldsymbol{x})$ and that has a much higher amplitude than the signal field. In an imaging system with an RA, the LO is the field that comes from the RA. In a system without an RA, a proportion of the excitation field serves as the LO. In the system without the RA, the delay time $\tau$ between the LO and the signal field is always zero, whereas in the system with the RA, $\tau$ is defined by the RA's pathlength. The LO interferes with the signal field on the photodetector to form the image, where the LO-to-signal (L-S) interference amplitude is much greater than the signal-to-signal (S-S) interference amplitude. Therefore, when an LO is present, the latter amplitude can be ignored.

Using the LO and the different light-matter interaction types, the imaging systems can be categorized into four types (*7*): category 1 (C1): coherent interaction with LO; C2: coherent interaction without LO; C3: incoherent interaction with LO; and C4: incoherent interaction without LO (see Method: Image formation formula). In C1, the signal field generated by the coherent interaction interferes with the LO. C3 is a special case in which the vacuum field applied to the incoherent interaction acts as the LO (*33*).

## Universal Image Formation Theory

### *1. Generalized image-formation formula in four-dimensional space-time domain*

In our theory, any imaging system can be characterized using its own instrumental function, which is defined using four components (see Method: Image formation formula): (1) the excitation system ASFs, (2) the interaction type, (3) the signal collection system ASF, and (4) the LO with delay time $\tau$ , if the LO exists. Components (1) and (2) determine the single-molecule polarization $\hat{p}^{(i)}(\boldsymbol{x},t)$ (see Method: Single-molecule polarization), component (3) is defined as a retarded Green's function (see Method: Retarded Green's function for the signal field), and additional issues related to component (4) are discussed in Method (about the LO).

The generalized expression for the image, $I(-\boldsymbol{x}',t_{\mathrm{d}},\tau)$, is a function of the 3D space coordinate $\boldsymbol{x}'$, the signal-field detection time $t_{\mathrm{d}}$, and the delay time $\tau$ between the signal field and the LO, and is written as:

$$I(-\boldsymbol{x}',t_{\mathrm{d}},\tau)=Lo(\,t_{\mathrm{d}}-\tau)+e^{i\theta}T^{*}(t_{\mathrm{d}}-\tau)\iiint N(\boldsymbol{x}'+\boldsymbol{x})h_{\mathrm{t}}(\boldsymbol{x},t_{\mathrm{d}})d^{3}\boldsymbol{x}+c.c.$$

$$+\left|\iiint N(\boldsymbol{x}'+\boldsymbol{x})h_{\mathrm{t}}'(\boldsymbol{x},t_{\mathrm{d}})\ d^{3}\boldsymbol{x}\right|^{2}, \tag{1}$$

where $h_{\mathrm{t}}(\boldsymbol{x},t_{\mathrm{d}})$ and $h_{\mathrm{t}}'(\boldsymbol{x},t_{\mathrm{d}})$ are the instrumental functions in the presence and absence of the LO, respectively. $Lo(t_{\mathrm{d}}-\tau)$ represents the LO intensity and $T(t_{\mathrm{d}}-\tau)$ corresponds to a complex amplitude of the LO. $t_{\mathrm{d}}-\tau$ represents the arrival time of the LO at the detector. $\theta$ is a phase offset that is explained later in this section. The derivation of Eq. (1) can be found in Methods.

Equation (1) [and the more accurate expression of Eq. (51) in Methods] is the universal image formation formula that covers all four categories defined above. The first and fourth terms represent the LO-to-LO (L-L) and S-S interferences, respectively, while the second and third terms represent the L-S interference and its complex conjugate, respectively. For systems in category C1, the coherent interaction with the LO, i.e., the fourth term of Eq. (1) (S-S interference) is significantly smaller than the other terms and thus is negligible. For categories C2 and C4, i.e., imaging systems without LOs, LO-related terms (the first to third terms) become zero. For category C3, involving incoherent interaction with the LO, the first term (L-L interference) cannot be observed because the LO is the vacuum field. In addition, similar to C1, the fourth term is negligibly small. Therefore, only the second and third terms contribute to the image formation.

To consider the phase offset $\theta$, category C1 is further divided into the transmission (C1-T) type without the RA and the reflection (C1-R) type with the RA. For C1-T, the phase shift $e^{i\theta}$ in the second term becomes $e^{i\pi/2}=i$ because of the Gouy phase shift, i.e., the phase difference between the signal field and the LO (*34*). Similarly, the phase shift of the third term becomes $-i$. For C1-R, the phase shifts in the second and third terms become $e^{-i\pi/2}=-i$ and $i$, respectively, based on consideration of the phase inversion of the LO caused by reflection from the reference mirror. Notably, however, C1-T includes holographic microscopy, which has an RA with LO phase shift $e^{-i\pi/2} = -i$. In category C3, the LO is the vacuum field that exists in all directions, and $e^{i\theta}$ is always $e^{i\pi/2} = i$, regardless of the transmission/reflection type, as shown in Method (Image formation formula: Category C3).

Note that although Eq. (1) is expressed using five variables, the two time-related variables $t_{\mathrm{d}}$ and $\tau$ are fourth arguments of the 4D space-time function. Specifically, for systems such as OCT, the exposure time is assumed to be sufficiently long, i.e., $I(-\boldsymbol{x}', t_{\mathrm{d}}, \tau)$ is integrated over $t_{\mathrm{d}}$ to give the 4D image $I(-\boldsymbol{x}',\ \tau)$. In contrast, in microscopy, usually $\tau = 0$ or the LO is absent, and the 4D image can be thus expressed using $I(-\boldsymbol{x}', t_{\mathrm{d}})$. Because both $t_{\mathrm{d}}$ and $\tau$ represent the time, the Fourier pair of both variables represents the light frequency $\nu$.

## *2. Optical resolution in four-dimensional frequency domain*

We move into the 4D frequency domain to discuss the 4D optical resolution, which consists of three spatial resolutions and a temporal resolution. Because of the strict definition of the 4D frequency cutoff, the optical resolution can be evaluated and compared with other imaging systems more clearly in the frequency domain than in the real-space domain.

We define the 4D aperture $A_4(\boldsymbol{f}, \nu/c)$ as the 4D Fourier transform of the instrumental functions $h_{\mathrm{t}}(\boldsymbol{x}, t_{\mathrm{d}})$ or $h_{\mathrm{t}}'(\boldsymbol{x}, t_{\mathrm{d}})$, where selection of $h_{\mathrm{t}}$ or $h_{\mathrm{t}}'$ is determined by the terms that are dominant in Eq. (1). Specifically, for systems with the LO (C1 and C3), $h_{\mathrm{t}}$ is selected, and for those without the LO (C2 and C4), $h_{\mathrm{t}}'$ is selected. In our theory, the instrumental function contains the light-matter interaction type, while the object to be observed is simply the molecular density $N(\boldsymbol{x})$. Note that in our previous theory (*7*), the object was a susceptibility distribution $\chi^{(i)}(\boldsymbol{x}) = N(\boldsymbol{x})\chi_{\mathrm{p}}^{(i)}$. For simplicity, we assume an infinitely fast detector-response speed here. Therefore, the temporal property of the instrumental function represents the temporal light-matter interaction properties, including the longitudinal and transverse relaxation times ($T_1$ and $T_2$, respectively) (*19, 21*). In other words, the 4D aperture also represents the spectroscopic properties of the light-matter interaction.

In practice, the 4D aperture can be computed readily in the frequency domain. The aperture can be constructed using $i$+1 4D pupil functions, i.e., Fourier transforms of the ASFs, in a specific imaging system with a $\chi^{(i)}$-derived interaction. The relevant Feynman diagram contains $i$ arrows that correspond to the 4D pupil functions for the excitation fields and, where necessary, the vacuum field. Additionally, a single 4D signal-collection pupil function always exists in the imaging system, i.e., an imaging system has $i$+1 pupil functions but it has only one 4D aperture. The relationship between the pupil functions and the arrows in a diagram are found in the correspondence rule, which is summarized in Table 1 in Method. Note that the signal-collection pupil forms part of the conic surface restricted by the signal collection system's NA (as exemplified in Fig. 4A) and accepts all positive light frequencies. The temporal property of the light-matter interaction is considered within the 4D aperture by using a complex Lorentzian function. Note also that the positive direction of $\boldsymbol{f}$ is redefined according to $\widetilde{N}(\boldsymbol{f})$ because the axis direction is flipped after the Fourier transform.

Here we present the practical steps taken to compute the 4D aperture.

(i) Select the excitation, vacuum, and signal collection pupil functions from Table 1. For example, Fig. 4A and 4B depict the pupils and the diagram, respectively, for confocal stimulated Raman gain (SRG) microscopy (*35*), which is a type of PP microscopy (*18*).

(ii) For each excitation pupil, compute a corresponding complex Lorentzian function, which is related to the relaxation time after the transition (*19, 21*). The SRG example is shown in Fig. 4B.

(iii) Combine the excitation pupils and Lorentzian functions as follows. The computation is performed in time order from the bottom of the diagram. In the first step, the first excitation pupil is multiplied by the first complex Lorentzian function $L_1(\nu)$. For the subsequent steps, the function resulting from the previous step is convolved with the next excitation pupil function in four dimensions and is then multiplied by the next Lorentzian $L_j(\nu)$, where $j$ is an excitation sequence number. This step is repeated until the end of the diagram is reached to provide $A_{\mathrm{ex}}(-\boldsymbol{f},\nu)$; the SRG example is shown in Fig. 4A.

(iv) Finally, the 4D aperture is computed by convolving the function $A_{\mathrm{ex}}(-\boldsymbol{f},\nu)$ that resulted from (iii) with the signal-collection pupil function in three dimensions $(f_x, f_y, f_z)$ at each light frequency. Note that the operation is three-dimensional for this convolution only, which stems from the integration over $t$ (see Method: Image formation formula). The 4D aperture for confocal SRG microscopy is depicted in Fig. 4C.

Two points should be noted here. First, Table 1 is exhaustive and is subdivided into four tables that correspond to categories C1–C4. Note here that for confocal imaging systems, including confocal SRG microscopy, this table can be simplified into Table 2, irrespective of the categories. Second, for imaging systems involving only nonresonant excitation, including most OCT, SHG, and THG microscopy systems, the effective portion of the complex Lorentzian function can be regarded as a real constant.

For reference, the 4D aperture of a confocal FL microscopy system is described in Fig. 4D and a temporal profile along the $t_{\mathrm{d}}$ axis calculated by Fourier transforming the 4D aperture is also depicted in Fig. 4E.

## Derivation of individual image formation formulas from universal equation

The Fourier components of an object $\widetilde{N}(\boldsymbol{f})$ spread over the 4D frequency space are uniformly distributed along the $\nu$ axis. An optical imaging system captures a specific region of these Fourier components by multiplying the image with the 4D aperture. To form a 3D image, this captured 4D spectrum is reduced to a 3D spectrum by integrating along one of the $f_z$ or $\nu$ axes. These integrations correspond to fixing $z'$ or $\tau$, respectively. The unfixed variable, i.e., $\tau$ or $z'$, then appears as one of the axes of the 3D image.

Here, we study four examples of the image formation types by using the universal image formation formula [Eq. (1)]. Image formation type 1 (IF1) is a $\tau$-scan system with an LO from the RA, e.g., OCT, where the sample stage $(z')$ is fixed and only the reference mirror is scanned along $\tau$. IF2 is a $z'$-scan system with an LO but without an RA; examples include FL, stimulated Raman scattering, PP, TPEF, and bright-field microscopy. IF3 is a $z'$-scan system without an LO, e.g., SHG, THG, and dark-field microscopy. In the IF2 and IF3 systems, the sample stage or the objective lens is usually scanned along $z'$. IF4 is a $(z',\tau)$-scan system with an LO from the RA, where both $z'$ and $\tau$ are scanned independently; examples include some interferometric and holographic microscopy techniques. These four imaging types cover almost all practical imaging systems, and Eq. (1) covers all far-field imaging systems.

Next, we formulate the image formation processes for IF1 to IF4. For IFs with an LO, i.e., IF1, IF2, and IF4, we introduce a quantity $B_4(\boldsymbol{f},\nu/c) = \widetilde{N}(\boldsymbol{f})\{\tilde{T}^*(\nu/c)A_4(\boldsymbol{f},\nu/c)\}$, which is the molecular density spectrum $\widetilde{N}(\boldsymbol{f})$ when windowed by the product of the complex conjugate of the LO spectrum $\tilde{T}^*(\nu/c)$ and the 4D aperture $A_4(\boldsymbol{f},\nu/c)$. Because $\tilde{T}^*(\nu/c)$ and $A_4(\boldsymbol{f},\nu/c)$ are amplitudes, their product is a squared amplitude, i.e., an intensity. This intensity nature eases intuitive depiction of the image formation process, as shown in Fig. 4A.

In type IF1, e.g., OCT, the image is $I_{\mathrm{IF1}}(-x',-y',\tau) = \int I_2(-\boldsymbol{x}',t_{\mathrm{d}},\tau)dt_{\mathrm{d}}|_{z'=0}$, where $I_2$ is the second term in Eq. (1). Here, the integration over $t_{\mathrm{d}}$ corresponds to a sufficiently long exposure time when compared with the oscillation of light. $z'=0$ indicates the absence of a $z'$-scan. The Fourier transform of the image is written as $\tilde{I}_{\mathrm{IF1}}(f_x,f_y,\nu/c) = -i\int B_4(\boldsymbol{f},\nu/c)df_z$, where $-i$ corresponds to $\theta = -\pi/2$ in Eq. (1), which originates from reflection from the reference mirror. This frequency-domain imaging representation is depicted schematically in Fig. 5A.

For OCT, time-domain (*37*) and Fourier domain (*38*) systems exhibit identical images with the same optical resolution if the experimental conditions are adjusted appropriately (see Supplementary Note 5). The most notable characteristic of OCT is that the image $I_{\mathrm{IF1}}(-x', -y', \tau)$ is not a function of $z'$ because the Fourier components of the object are integrated over $f_z$ in the frequency domain, as shown in Fig. 5A. Because the object's spatial frequency component is in the $(f_x, f_y, f_z)$-space, this integration causes a partial information loss, particularly in significantly out-of-focus regions (*39*), if the 4D aperture is thick in the $f_z$ direction. Our 4D formulation and depiction in Fig. 5A imply that if the 4D aperture is sufficiently thin, then an OCT image frequency, which is in $(f_x, f_y, \nu/c)$-space, can be back-projected perfectly into the $(f_x, f_y, f_z)$-space, as shown in Fig. 5B. This thin aperture can be achieved by using plane wave illumination, i.e., by using an excitation NA of zero. This depiction will enable comprehensive interpretation of several OCT-based refocusing and aberration correction methods.

In IF2, because the LO exists, the second and third terms of Eq. (1) ($I_2$ and $I_3$, respectively) dominate the image. In addition, because the RA does not exist and the LO is a proportion of the excitation light, the LO delay $\tau$ is always zero. Therefore, the image becomes $I_{\mathrm{IF2}}(-x', -y', -z') = \int [I_2(-\boldsymbol{x}', t_{\mathrm{d}}, \tau = 0) + I_3(-\boldsymbol{x}', t_{\mathrm{d}}, \tau = 0)] dt_{\mathrm{d}}$. Here, the Fourier components of the object $\tilde{N}(\boldsymbol{f})$ captured via the 4D aperture are integrated over $\nu$, as shown in Fig. 5A. Note that $\tilde{N}(\boldsymbol{f})$ is uniform in the $\nu$ direction. The Fourier transform of the image is $\tilde{I}_{\mathrm{IF2}}(f_x, f_y, f_z) = e^{i\theta} \int B_4(\boldsymbol{f}, \nu/c) d\nu + c.c.$, where $e^{i\theta}$ becomes $i$.

In IF3, because the LO is absent, the image consists of only the fourth term from Eq. (1), $I_4$, in the form $I_{\mathrm{IF3}}(-x', -y', -z') = \int I_4(-\boldsymbol{x}', t_{\mathrm{d}}, 0) dt_{\mathrm{d}}$. Using the 4D aperture, the Fourier transform of the image is given by $\tilde{I}_{\mathrm{IF3}}(f_x, f_y, f_y) = \mathrm{AC}[\tilde{N}(\boldsymbol{f}) A_4(\boldsymbol{f}, \nu/c)]\big|_{\nu=0}$, where $\mathrm{AC}[\cdots]$ denotes autocorrelation.

IF4 is the final example and includes interferometric and holographic microscopy techniques. These imaging modalities have an LO coming from the RA (*30, 31*), and both $z'$and $\tau$ are scanned independently. This type of the system initially acquires a 4D meta-image as $I_{\mathrm{IF4}}(-\boldsymbol{x}', \tau) = \int I_2(-\boldsymbol{x}', t_{\mathrm{d}}, \tau) dt_{\mathrm{d}}$, where only the second term of Eq. (1) is extracted by using the interference. The final 3D image is then acquired by setting $\tau = 0$ such that $I_{\mathrm{IF4}}(-x', -y', -z', 0)$. This dimensionality reduction from four to three is depicted in the frequency domain, as illustrated in Fig. 5A.

Because a 4D meta-image $I_{\mathrm{IF4}}(-\boldsymbol{x}', \tau)$ can be obtained, an arbitrary 3D image such as a microscopy image $I_{\mathrm{IF4}}(-x', -y', -z', 0)$ or an OCT image $I_{\mathrm{IF1}}(-x', -y', 0, \tau)$ can be selected, as depicted in Fig. 5A. The former is a spatially 3D microscopic image similar to IF2, but it allows both amplitude and phase information to be obtained. The latter is a 3D image that is identical to that of IF1. The image frequency of $I_{\mathrm{IF4}}(-\boldsymbol{x}', 0)$ is found in the form $\tilde{I}_{\mathrm{IF4}}(f_x, f_y, f_z) = e^{i\theta} \int B_4(\boldsymbol{f}, \nu/c) d\nu$, where $e^{i\theta}$ becomes $-i$ because of the mirror reflection in the RA. Unlike IF1, the image frequency in this case is provided by integrating $B_4(\boldsymbol{f}, \nu/c)$ over $\nu$ rather than $f_z$. Therefore, the object frequency within the 4D aperture remains without information loss.

## Discussion

In this work, we constructed the 4D image formation theory by introducing the 4D pupil function on the surface of the light cone in the 4D frequency space and defining the 4D aperture as the instrumental function. The 4D aperture can be an indicator of the optical resolution limit and it is calculated from *i*+1 4D pupil functions, where *i* represents the interaction order. Using our theoretical framework, almost all optical imaging systems can be associated with each other and precisely understood the relationship, which allows for the connection between the experts working with different types of imaging systems. Although some image formation types exist, the image formation formula of each imaging system can be derived from the 4D aperture.

To evaluate the maximum value of the frequency cutoff, we applied first Born approximation, which does not include the effect of multiple scattering that causes the optical resolution to be worse. Since the magnetic field does not strongly interact with the molecules in a sample if the major target is a biomedical specimen, we treated only the electric field. We showed that the integration over $\omega$ in the 4D frequency space is related to the optical microscopy, while the integration over $f_z$ is associated with OCT. Our theory can provide the method to convert an OCT image into the microcopy image that obediently reflects the sample structure.

## Methods

### Operators in the space-time domain

The operators in the space-time domain $\hat{a}(x,y,z,t)$ can be converted from those in the wavevector domain $\hat{a}(\boldsymbol{k})$ as follows:

$$\hat{a}(x,y,z,t) = \frac{\sqrt{\hbar c^2}}{\sqrt{(2\pi)^3}} \int \frac{1}{\sqrt{2\omega}} \hat{a}(\boldsymbol{k}) e^{i(k_x x + k_y y + k_z z - \omega t)}\, d^3\boldsymbol{k}$$
$$= \sqrt{\hbar c^2} \int \frac{1}{\sqrt{2\nu}} \hat{a}(\boldsymbol{f}) e^{i2\pi(f_x x + f_y y + f_z z - \nu t)}\, d^3\boldsymbol{f}, \quad (2)$$

where $\hbar$ is the reduced Planck constant (*27*). Hereafter, we use the natural units: $\hbar = 1$ and $c = 1$.

### 4D Fourier optics

We select an optical axis oriented in the $z$ direction. In the laser excitation system, $(x', y')$ corresponds to the amount of a laser shift or a sample stage displacement, whereas in the Kohler illumination system, $(x', y')$ corresponds to a position on the image plane. In both systems, a $z'$-scan is performed by varying the relative distance between the sample and the objective lens.

To address all light-matter interactions, including fluorescence, within the same framework, we expand double-sided Feynman diagrams that describe how the density matrix of the molecular ensemble evolves over time through interaction with light beams (electric fields) (*19*). In the diagram for incoherent interactions such as fluorescence, we add the vacuum field as one of the possible de-excitation fields. All linear, nonlinear, coherent, and incoherent interactions can be described using our diagram, which includes some arrows to represent the electric fields, as shown in Fig. 2B and 2C. To incorporate the vacuum field, we use the annihilation operators $\hat{E}(x,y,z,t)$, which are defined as

$$\hat{E}(x,y,z,t) = \int \frac{1}{\sqrt{2\nu}} \hat{E}(\boldsymbol{f}) e^{i2\pi(f_x x + f_y y + f_z z - \nu t)}\, d^3\boldsymbol{f}, \quad (3)$$

and the creation operators $\hat{E}^{\dagger}(x,y,z,t)$ for all electric fields acting as excitation fields, including the vacuum field if necessary and the signal field; these operators are 4D operators in space-time $(x,y,z,t)$. This 4D time-space domain relates to the 4D frequency domain $(f_x, f_y, f_z, \nu)$ via the 4D Fourier transform:

$$\tilde{f}(f_x, f_y, f_z, \nu) = \int f(x,y,z,t) e^{i2\pi(f_x x + f_y y + f_z z - \nu t)}\, dxdydzdt. \quad (4)$$

## Quantum Liouville equation

To formulate a universal image formation theory that can treat all interactions, we begin with the quantum Liouville equation (*19*):

$$\frac{\partial}{\partial t}\rho(t) = \frac{1}{i\hbar}[H(t), \rho(t)], \tag{5}$$

where $\rho(t)$ represents the time evolution of the density matrix and $H(t)$ represents an interaction Hamiltonian. Here, $\rho(t)$ and $H(t)$ are represented in the interaction picture. Equation (5) can be solved by performing a perturbation expansion as shown below:

$$\rho(t) = \rho^{(0)}(t_0) + \sum_{n=1}^{\infty} \rho^{(n)}(t), \tag{6}$$

$$\rho^{(n)}(t) = \left(\frac{1}{i\hbar}\right)^n \int_{t_0}^{t} dt_n \cdots \int_{t_0}^{t_3} dt_2 \int_{t_0}^{t_2} dt_1 \left[H(t_n), \cdots [H(t_2), [H(t_1), \rho^{(0)}(t_0)]]\right], \tag{7}$$

where *n* is the interaction order and $-\infty = t_0 < t_1 < t_2 \cdots < t_n < t$ are times. The solution to the Liouville equation for $\rho(\boldsymbol{x}, t)$ can be expanded to multiple terms and each of these terms corresponds to a Feynman diagram. Hereafter, we use the density matrix in the Schrodinger picture $\rho_{\mathrm{S}}(t)$ for the imaging systems. The relationship between the interaction and Schrodinger pictures is given by:

$$\rho(t) = e^{+iH_0 t/\hbar} \rho_{\mathrm{S}}(t) e^{-iH_0 t/\hbar} \tag{8}$$

$$H(t) = e^{+iH_0 t/\hbar} H_{\mathrm{S}}(t) e^{-iH_0 t/\hbar} \tag{9}$$

where $H_0$ is an unperturbed Hamiltonian. To apply the quantum Liouville equation to imaging systems, we add the spatial coordinate $\boldsymbol{x} = (x, y, z)$ by considering lens effects, i.e., we obtain $\rho_{\mathrm{S}}(\boldsymbol{x}, t)$.

The expectation value for the detected signal is expressed as

$$I(\boldsymbol{x}', \tau) = \int \left| -iL(t_d - \tau) + \iint_{x\,t} N(\boldsymbol{x}' - \boldsymbol{x}) \mathrm{tr}\big(\mu \rho_{\mathrm{S}}(\boldsymbol{x}, t)\big) h_{\mathrm{col}}(-\boldsymbol{x}, t_d - t) d^3\boldsymbol{x} dt \right|^2 dt_d, \tag{10}$$

where

- $\tau$: Delay time in time-domain OCT (TD-OCT; zero in microscopy);
- $t_{\mathrm{d}}$ : Detection time;
- $L(t_{\mathrm{d}} - \tau)$: Light that is not influenced by an object;
- $\boldsymbol{x}$: Object coordinate;
- $t$: Signal radiation time from a molecule;
- $\boldsymbol{x}'$ : Laser beam scanning displacement;
- $\mu$: Dipole moment of a target molecule (matrix).
- $N(\boldsymbol{x})$: Density distribution of the target molecules (object).
- $h_{\mathrm{col}}(-\boldsymbol{x}, t_{\mathrm{d}} - t)$ : Green's function of the signal collection system.

In Eq. (10), $\mathrm{tr}\big(\mu\rho_{\mathrm{S}}(\boldsymbol{x}, t)\big)$ contains all interactions, but we can usually select one type from these interactions by devising a suitable experimental setup. After selecting a single diagram, we then express the signal intensity as

$$I(-\boldsymbol{x}', \tau) = \int \left| -iL(t_d - \tau) + \int N(\boldsymbol{x}' + \boldsymbol{x}) h_{\mathrm{t}}(\boldsymbol{x}, t_d) d^3\boldsymbol{x} \right|^2 dt_d$$

$$\begin{cases} \approx \mathrm{Const.} + i \displaystyle\int N(\boldsymbol{x}' - \boldsymbol{x}) \{L^*(-\tau) \otimes^t h_{\mathrm{t}}(\boldsymbol{x}, \tau)\} d^3\boldsymbol{x} + \mathrm{c.c.} \quad \text{(with LO)} & (11) \\ = \displaystyle\int \left| \int N(\boldsymbol{x}' - \boldsymbol{x}) h_{\mathrm{t}}(\boldsymbol{x}, t_d) d^3\boldsymbol{x} \right|^2 dt_d \Big|_{=0} \quad \text{(without LO)} & (12) \end{cases}$$

where $\otimes^t$ denotes a convolution with respect to the time and $h_t(\boldsymbol{x}, t_d)$ represents an instrumental function of the imaging system with the relevant interaction, which is calculated by performing the integral over $t$ after selecting a specific diagram:

$$h_t(\boldsymbol{x}, t_d) = \int Dia(\boldsymbol{x}, t) h_{\text{col}}(-\boldsymbol{x}, t_d - t) dt. \tag{13}$$

where $Dia(\boldsymbol{x}, t)$ represents the relevant term in $\text{tr}\big(\mu\rho_S(\boldsymbol{x}, t)\big)$. The instrumental function $h_t(\boldsymbol{x}, t')$, in which $t' = \tau$ or $t_d$, determines the resolution limit and is dependent on the type of interaction. Note that we consider a 4D meta image here to unify the image formation theory.

## Excitation pupil function

### *1. Laser excitation*

We consider the excitation field produced by a laser source. The coherent state $|\alpha_f\rangle$ is generated using the displacement operator $\widehat{D}(\alpha_f) = e^{-|\alpha_f|^2/2} e^{\alpha_f \hat{a}^\dagger} e^{-\alpha_f^* \hat{a}}$ as follows: $|\alpha_f\rangle = \widehat{D}(\alpha_f)|0\rangle$ (*40*), where $|0\rangle$ denotes the vacuum state composed of all modes (wavenumber $\boldsymbol{k} = 2\pi\boldsymbol{f}$):

$$|0\rangle = \prod_{\boldsymbol{f}} |0\rangle_{\boldsymbol{f}}. \tag{14}$$

We define the annihilation operator for the excitation field in the object $\hat{a}_{\text{EX}}(\boldsymbol{x}, t)$ as:

$$\hat{a}_{\text{EX}}(\boldsymbol{x}, t) = \int \frac{1}{2\nu} p_{\text{ex}}(\nu) NA_{\text{ex}}(\boldsymbol{f}, \nu) \widehat{D}^\dagger(\alpha_f) \hat{a}(\boldsymbol{f}) \widehat{D}(\alpha_f) e^{i2\pi(\boldsymbol{f}\cdot\boldsymbol{x} - \nu t)}\, d^3\boldsymbol{f}, \tag{15}$$

where a 4D function $NA_{\text{ex}}(\boldsymbol{f}, \nu)$ determines the NA and the transmittance of the excitation system that might contain aberrations, and $p_{\text{ex}}(\nu)$ denotes the normalized spectrum of the laser source. The ASF formed by the excitation field in the object is then given by

$$\begin{aligned} ASF_{\text{ex}}(\boldsymbol{x}, t) &= \Theta(t)\langle 0|\hat{a}_{\text{EX}}(\boldsymbol{x}, t)|0\rangle \\ &= \int \frac{1}{2\nu} \delta(|\boldsymbol{f}| - \nu)\, \alpha_f\, p_{\text{ex}}(\nu) NA_{\text{ex}}(\boldsymbol{f}, \nu)\; e^{i2\pi(\boldsymbol{f}\cdot\boldsymbol{x} - \nu t)}\, d^3\boldsymbol{f} d\nu \\ &= \int P_{\text{ex}}(\boldsymbol{f}, \nu)\, e^{i2\pi(\boldsymbol{f}\cdot\boldsymbol{x} - \nu t)}\, d^3\boldsymbol{f} d\nu, \end{aligned} \tag{16}$$

where

$$P_{\text{ex}}(\boldsymbol{f}, \nu) = \frac{1}{2\nu} \delta(|\boldsymbol{f}| - \nu) \alpha_f\, p_{\text{ex}}(\nu) NA_{\text{ex}}(\boldsymbol{f}, \nu), \tag{17}$$

where $\Theta(t)$ is the Heaviside step function that satisfies the following expression:

$$\Theta(t) = \int \delta_+(\nu) e^{-i2\pi\nu(t)} d\nu, \tag{18}$$

where

$$\delta_+(|\boldsymbol{f}| - \nu) = \frac{1}{2}\delta(|\boldsymbol{f}| - \nu) + \frac{1}{i2\pi(|\boldsymbol{f}| - \nu)}. \tag{19}$$

$P_{\text{ex}}(\boldsymbol{f}, \nu)$ corresponds to one of the 4D pupil functions in the excitation system, and $\alpha_f\, p_{\text{ex}}(\nu) = S_{\text{ex}}(\nu)$ represents the light source spectrum. For convenience, we maintain the ASF as the operator $ASF_{\text{ex}}(\boldsymbol{x}, t) \to \hat{a}_{\text{ex}}(\boldsymbol{x}, t)$:

$$\hat{a}_{\text{ex}}(\boldsymbol{x}, t) = \int \frac{1}{\alpha_f} P_{\text{ex}}(\boldsymbol{f}, \nu) \widehat{D}^\dagger(\alpha_f) \hat{a}(\boldsymbol{f}) \widehat{D}(\alpha_f) e^{i2\pi(\boldsymbol{f}\cdot\boldsymbol{x} - \nu t)}\, d^3\boldsymbol{f} d\nu, \tag{20}$$

where the relationship $\widehat{D}^\dagger(\alpha_f)\hat{a}(\boldsymbol{f})\widehat{D}(\alpha_f) = \hat{a}(\boldsymbol{f}) + \alpha_f$ is useful.

*2. Kohler illumination*

Similarly, we assume an operator for the Kohler illumination (*41*). We propose that the Kohler illumination state $|c_{\boldsymbol{f}}\rangle$ can be produced by an operator $\hat{C}(c_{\boldsymbol{f}})$ as follows: $|c_{\boldsymbol{f}}\rangle = \hat{C}(c_{\boldsymbol{f}})|0\rangle$. Here, we assume that any state can be expressed as a sum of coherent states:

$$|c_{\boldsymbol{f}}\rangle = \int c(\alpha_{\boldsymbol{f}})|\alpha_{\boldsymbol{f}}\rangle\, d^2\alpha_{\boldsymbol{f}} = \int c(\alpha_{\boldsymbol{f}})\hat{D}(\alpha_{\boldsymbol{f}})|0\rangle\, d^2\alpha_{\boldsymbol{f}} = \hat{C}(c_{\boldsymbol{f}})|0\rangle, \tag{21}$$

where $c(\alpha_{\boldsymbol{f}})$ is a coefficient and $\alpha_{\boldsymbol{f}}$ is a complex number. Using this assumption, the annihilation operator can then be written as:

$$\hat{a}_{\text{ill}}(\boldsymbol{x},t) = \int \frac{1}{C_{\boldsymbol{f}}} P_{\text{ill}}(\boldsymbol{f},\nu)\hat{C}^{\dagger}(c_{\boldsymbol{f}})\hat{C}(c_{\boldsymbol{f}})e^{i2\pi(\boldsymbol{f}\cdot\boldsymbol{x}-\nu t)}\, d^3\boldsymbol{f}d\nu, \tag{22}$$

where

$$P_{\text{ill}}(\boldsymbol{f},\nu) = \frac{1}{2\nu}\delta(|\boldsymbol{f}|-\nu)R(\boldsymbol{f})S_{\text{ill}}(\nu)NA_{\text{ex}}(\boldsymbol{f},\nu) = R(\boldsymbol{f})S_{\text{ill}}(\nu)\mathbb{P}_{\text{ill}}(\boldsymbol{f},\nu), \tag{23}$$

and $P_{\text{ill}}(\boldsymbol{f},\nu)$ corresponds to the 4D pupil function of the Kohler illumination system; $R(\boldsymbol{f})$ represents a function with a random phase that describes the illumination system characteristics, $S_{\text{ill}}(\nu)$ denotes the amplitude spectrum of the light source, $\mathbb{P}_{\text{ex}}(\boldsymbol{f},\nu) = (1/2\nu)\delta(|\boldsymbol{f}|-\nu)NA_{\text{ex}}(\boldsymbol{f},\nu)$, and $C_{\boldsymbol{f}} = \langle c_{\boldsymbol{f}}|c_{\boldsymbol{f}}\rangle$. The relationship between the operator and the 4D pupil function can then be determined as follows:

$$\langle 0|\hat{a}_{\text{ill}}(\boldsymbol{x},t)|0\rangle = \int P_{\text{ill}}(\boldsymbol{f},\nu)e^{i2\pi(\boldsymbol{f}\cdot\boldsymbol{x}-\nu t)}d^3\boldsymbol{f}d\nu \equiv K(\boldsymbol{x},t), \tag{34}$$

where $K(\boldsymbol{x},t)$ represents the light amplitude distribution in the object. We also obtain the following relationship:

$$K(\boldsymbol{x}_1,t_1)K^*(\boldsymbol{x}_2,t_2) = \Gamma_{12}(\boldsymbol{x}_1-\boldsymbol{x}_2,t_1-t_2), \tag{35}$$

where $\Gamma_{12}(\boldsymbol{x},t)$ denotes the mutual intensity (*36*).

## Single-molecule polarization

*i* excitation fields interact with the molecules before the signal field is emitted, and therefore we must consider the single-molecule polarization $\hat{p}^{(i)}(\boldsymbol{x},t)$, which includes all excitation fields, e.g., $\hat{a}^{\dagger}_{\text{ex1}}(\boldsymbol{x},t)$, $\hat{a}_{\text{ex2}}(\boldsymbol{x},t)$, $\hat{a}_{\text{ex3}}(\boldsymbol{x},t)$, and so on. On the basis of a nonlinear optics approach (*19*), $\hat{p}^{(i)}(\boldsymbol{x},t)$, which varies depending on the specific diagram, can be calculated using the relevant diagram. We take the SRG case as an example and use it to explain the calculation rule, but all interactions occur in the same manner. The single-molecule polarization $\hat{p}^{(i)}(\boldsymbol{x},t)$ can be calculated as follows (*19, 21, 42*):

$$\begin{aligned}\hat{p}^{(i)}_{\text{SRG}}(\boldsymbol{x},t) = \mu_{\text{bc}}e^{i\omega_{\text{bc}}t} \\ \times\frac{i\mu_{\text{cg}}}{\hbar}\int_{-\infty}^{t} e^{i\omega_{\text{cg}}t_3}\,\hat{a}_{\text{ex3}}(\boldsymbol{x},t_3-t_3') \\ \times\frac{-i\mu_{\text{ab}}}{\hbar}\int_{-\infty}^{t_3} e^{i\omega_{\text{ab}}t_2}\,\hat{a}_{\text{ex2}}(\boldsymbol{x},t_2-t_2') \\ \times\frac{-i\mu_{\text{ga}}}{\hbar}\int_{-\infty}^{t_2} e^{i\omega_{\text{ga}}t_1}\,\hat{a}^{+}_{\text{ex1}}(\boldsymbol{x},t_1-t_1')dt_1dt_2\,dt_3,\end{aligned} \tag{36}$$

where $t_1'$, $t_2'$, and $t_3'$ are the central times for the first, second, and third excitation laser pulses, respectively, $\omega_{\text{bc}} = \omega_{\text{b}}-\omega_{\text{c}}$ denotes the difference between the two energy levels b and c in the molecule, and $\mu_{\text{bc}}$ represents the transition dipole moment from energy level c to energy level b.

The time-order relationship is $t_1' \leq t_2' \leq t_3' \leq t < t_\mathrm{d}$. The calculation can be performed by following the rule below.

Calculation rule for $\hat{p}^{(i)}(\boldsymbol{x}, t)$:

R1. Multiple integrals based on the number of the excitation fields are performed in time order. Because each integral includes the relevant excitation-field operator, these integrals are arranged in order from the right.

R2. In the Feynman diagram of interest (see e.g. Fig. 4B), arrows corresponding to each excitation field are shown. The energy levels above and below each arrow determine the energy differences in the integral (e.g., $\omega_\mathrm{ga}$ for the first excitation) and the transition dipole moment (e.g., $\mu_\mathrm{ga}$ for the first excitation). The subscripts "ga" for both parameters are identical and are arranged in order from the right, but the appropriate counting order is upward if the arrow is located on the left side and downward if the arrow is located on the right side.

R3. The top of the diagram after all excitations represents the state of the molecules when radiating the signal. The terms for the energy difference and the transition dipole moment $\mu_\mathrm{bc} e^{i\omega_\mathrm{bc} t}$ for the signal are multiplied. The subscripts are arranged in order from the right, where the counting order is from left to right. Therefore, all subscripts, including those of the excitations, are counted clockwise.

R4. If the arrow for the excitation is located on the left side, then $i/\hbar$ is multiplied, but if the arrow is located on the right side, $-i/\hbar$ is multiplied.

In the case of a continuous excitation field, the polarization at $\boldsymbol{x}$ can be expressed simply as $\varepsilon_0 \chi_\mathrm{p}^{(i)} \hat{E}_\mathrm{ex}^{(i)}(\boldsymbol{x}, t)$, where $\varepsilon_0$ is the vacuum permittivity and $\hat{E}_\mathrm{ex}^{(i)}(\boldsymbol{x}, t)$ corresponds to the product of all operators for the excitation fields, e.g., $\hat{a}_\mathrm{ex1}(\boldsymbol{x}, ct)$, in the relevant diagram (*19*, *21*).

## Retarded Green's function for the signal field

We consider a retarded Green's function for the signal field (*24*) that radiates from a single point in the object and propagates toward the detector through the signal collection system. Because the signal photon is created at a space-time $(\boldsymbol{x}, t)$ and is annihilated at $(\boldsymbol{x}_\mathrm{d}, t_\mathrm{d})$, the Green's function can then be written as

$$\begin{aligned} G_\mathrm{ret}(\boldsymbol{x}_\mathrm{d} - \boldsymbol{x}, t_\mathrm{d} - t) &= \Theta(t_\mathrm{d} - t)\langle 0|\hat{a}_\mathrm{col}(\boldsymbol{x}_\mathrm{d}, t_\mathrm{d})\hat{a}_\mathrm{sig}^{+}(\boldsymbol{x}, t)|0\rangle \\ &= \iint P_\mathrm{col}(\boldsymbol{f}, \nu) e^{i2\pi\{\boldsymbol{f}\cdot(\boldsymbol{x}_\mathrm{d} - \boldsymbol{x}) - \nu(t_\mathrm{d} - t)\}} d^3\boldsymbol{f} d\nu \\ &\equiv ASF_\mathrm{col}(\boldsymbol{x}_\mathrm{d} - \boldsymbol{x}, t_\mathrm{d} - t), \end{aligned} \tag{37}$$

with

$$\hat{a}_\mathrm{col}(\boldsymbol{x}_\mathrm{d}, t_\mathrm{d}) = \int \frac{1}{\sqrt{2\nu_d}} NA_\mathrm{col}(\boldsymbol{f}_\mathrm{d}, \nu_\mathrm{d}) \hat{a}(\boldsymbol{f}_\mathrm{d}) e^{i2\pi(\boldsymbol{f}_\mathrm{d}\cdot \boldsymbol{x}_\mathrm{d} - \nu_\mathrm{d} t_\mathrm{d})}\, d^3\boldsymbol{f}_\mathrm{d}, \tag{38}$$

$$\hat{a}_\mathrm{sig}^{+}(\boldsymbol{x}, t) = \int \frac{1}{\sqrt{2\nu}} \hat{a}^{+}(\boldsymbol{f}) e^{-i2\pi(\boldsymbol{f}\cdot \boldsymbol{x} - \nu t)}\, d^3\boldsymbol{f}, \tag{39}$$

$$P_\mathrm{col}(\boldsymbol{f}, \nu) = \frac{1}{2\nu} \delta(|\boldsymbol{f}| - \nu)\, NA_\mathrm{col}(\boldsymbol{f}, \nu), \tag{40}$$

where the 4D function $NA_\mathrm{col}(\boldsymbol{f}_\mathrm{d}, \nu_\mathrm{d})$ restricts the NA of the signal collection system and includes both the aberration and the transmittance. Note that $NA_\mathrm{col}(\boldsymbol{f}_\mathrm{d}, \nu_\mathrm{d})$ covers all positive light frequencies. $P_\mathrm{col}(\boldsymbol{f}, \nu)$ represents the 4D pupil function of the signal collection system.

## Pupil function for the LO from the laser source

We consider the LO signal propagating from one of the excitation laser sources toward the detector. In the same manner used for the excitation field, the ASF formed by the LO at the detector can be written as an operator:

$$\hat{a}_{\mathrm{lo}}(\boldsymbol{x}_{\mathrm{d}},t_{\mathrm{r}})=\int\frac{1}{2\nu_d}\delta(|\boldsymbol{f}_{\mathrm{d}}|-\nu_{\mathrm{d}})p_{\mathrm{lo}}(\nu_{\mathrm{d}})NA_{\mathrm{lo}}(\boldsymbol{f}_{\mathrm{d}},\nu_{\mathrm{d}})\hat{D}^{\dagger}(\alpha'_{\boldsymbol{f}_{\mathrm{d}}})\hat{a}(\boldsymbol{f}_{\mathrm{d}})\hat{D}(\alpha'_{\boldsymbol{f}_{\mathrm{d}}})\,e^{i2\pi(\boldsymbol{f}_{\mathrm{d}}\cdot\boldsymbol{x}_{\mathrm{d}}-\nu_{\mathrm{d}}t_{\mathrm{r}})}d^3\boldsymbol{f}_{\mathrm{d}}d\nu_{\mathrm{d}},\quad(41)$$

where $p_{\mathrm{lo}}(\nu_{\mathrm{d}})$ represents the normalized spectrum of the LO, $NA_{\mathrm{lo}}(\boldsymbol{f}_{\mathrm{d}},\nu_{\mathrm{d}})$ represents the NA of the LO, as determined by the LO system that may consist of the excitation/signal collection systems or by the RA, depending on the imaging system; $\alpha'_{\boldsymbol{f}_{\mathrm{d}}}$ denotes the average amplitude of the LO signal, and $t_{\mathrm{r}}=t_{\mathrm{d}}-\tau$. The 4D pupil function of the LO is given by

$$P_{\mathrm{lo}}(\boldsymbol{f}_{\mathrm{d}},\nu_{\mathrm{d}})=\frac{1}{2\nu_{\mathrm{d}}}\delta(|\boldsymbol{f}_{\mathrm{d}}|-\nu_{\mathrm{d}})\underbrace{\alpha'_{\boldsymbol{f}_{\mathrm{d}}}\,p_{\mathrm{lo}}(\nu_{\mathrm{d}})}_{S_{\mathrm{lo}}(\nu_{\mathrm{d}})}NA_{\mathrm{lo}}(\boldsymbol{f}_{\mathrm{d}},\nu_{\mathrm{d}})=S_{\mathrm{lo}}(\nu_{\mathrm{d}})\mathbb{P}_{\mathrm{lo}}(\boldsymbol{f}_{\mathrm{d}},\nu_{\mathrm{d}}),\quad(42)$$

where $S_{\mathrm{lo}}(\nu_{\mathrm{d}})=\alpha'_{\boldsymbol{f}_{\mathrm{d}}}\,p_{\mathrm{lo}}(\nu_{\mathrm{d}})$ represents the LO spectrum, and $\mathbb{P}_{\mathrm{lo}}(\boldsymbol{f}_{\mathrm{d}},\nu_{\mathrm{d}})$ denotes the signal collection pupil function when multiplied by the excitation pupil function over all light frequencies, which is shaped like part of the cone cut by the NA. The amplitude of the LO $L(\boldsymbol{x}_{\mathrm{d}},t_{\mathrm{r}})$ is given by:

$$\langle 0|\hat{a}_{\mathrm{lo}}(\boldsymbol{x}_{\mathrm{d}},t_{\mathrm{r}})|0\rangle=L(\boldsymbol{x}_{\mathrm{d}},t_{\mathrm{r}}).\quad(43)$$

## The vacuum field as the excitation field and LO

The annihilation operator of the vacuum field (*19*, *26*) in the object $\hat{a}_{\mathrm{vac}}(\boldsymbol{x},t)$ is given by:

$$\begin{aligned}\hat{a}_{\mathrm{vac}}(\boldsymbol{x},t)&=\iint\frac{1}{2\nu}\,\delta(|\boldsymbol{f}|-\nu)V(\boldsymbol{f})\hat{a}(\boldsymbol{f})e^{i2\pi\{\boldsymbol{f}\cdot\boldsymbol{x}-\nu t\}}d^3\boldsymbol{f}d\nu\\&=\int\frac{1}{2|\boldsymbol{f}|}\,V(\boldsymbol{f})\hat{a}(\boldsymbol{f})e^{i2\pi\{\boldsymbol{f}\cdot\boldsymbol{x}-|\boldsymbol{f}|t\}}d^3\boldsymbol{f},\end{aligned}\quad(44)$$

where $V(\boldsymbol{f})$ is a random phase function with a uniform modulus that describes the vacuum-field characteristics. Note that $V(\boldsymbol{f})$ is defined over all modes for all light frequencies. For convenience, we express $V(\boldsymbol{f},\nu)$ here as

$$V(\boldsymbol{f},\nu)=\frac{1}{2\nu}\,\delta(|\boldsymbol{f}|-\nu)V(\boldsymbol{f}).\quad(45)$$

In more detailed form, the phase of $V(\boldsymbol{f},\nu)$ is a random variable, and $V(\boldsymbol{f}_1,\nu_1)$ and $V(\boldsymbol{f}_2,\nu_2)$ are uncorrelated if $\boldsymbol{f}_1\neq\boldsymbol{f}_2$ and $\nu_1\neq\nu_2$, and the Fourier transform of $V(\boldsymbol{f},\nu)$ then becomes a function $\tilde{V}(\boldsymbol{x},t)$ that has similar characteristics. $\tilde{V}(\boldsymbol{x},t)$ also expresses a random phase function with a uniform modulus, the phase of $\tilde{V}(\boldsymbol{x},t)$ is a random variable, and $\tilde{V}(\boldsymbol{x}_1,t_1)$ and $\tilde{V}(\boldsymbol{x}_2,t_2)$ are uncorrelated if $\boldsymbol{x}_1\neq\boldsymbol{x}_2$ and $t_1\neq t_2$, which can be deduced by calculating an expectation value for the vacuum field.

We also consider the vacuum field as the LO at the detector plane as follows:

$$\begin{aligned}\hat{a}_{\mathrm{lo(v)}}(\boldsymbol{x}_{\mathrm{d}},t_{\mathrm{d}})&=\iint\frac{1}{2\nu_d}\,\delta(|\boldsymbol{f}_{\mathrm{d}}|-\nu_d)V(\boldsymbol{f}_{\mathrm{d}})NA_{\mathrm{col}}(\boldsymbol{f},\nu)\hat{a}(\boldsymbol{f}_{\mathrm{d}})e^{i2\pi\{\boldsymbol{f}_{\mathrm{d}}\cdot\boldsymbol{x}_{\mathrm{d}}-\nu_d t_{\mathrm{d}}\}}d^3\boldsymbol{f}_{\mathrm{d}}d\nu_d\\&=\int\frac{1}{2|\boldsymbol{f}_{\mathrm{d}}|}\,V(\boldsymbol{f}_{\mathrm{d}})NA_{\mathrm{col}}(\boldsymbol{f},|\boldsymbol{f}_{\mathrm{d}}|)\hat{a}(\boldsymbol{f}_{\mathrm{d}})e^{i2\pi\{\boldsymbol{f}_{\mathrm{d}}\cdot\boldsymbol{x}_{\mathrm{d}}-|\boldsymbol{f}_{\mathrm{d}}|t_{\mathrm{d}}\}}d^3\boldsymbol{f}_{\mathrm{d}}.\end{aligned}\quad(46)$$

By considering the time order $t_{\mathrm{d}}>t$, the calculation shown below can be performed:

$$\begin{aligned}\Theta(t_{\mathrm{d}}-t)\langle 0|\hat{a}_{\mathrm{vac}}(\boldsymbol{x},t)\hat{a}^{\dagger}_{\mathrm{lo(v)}}(\boldsymbol{x}_{\mathrm{d}},t_{\mathrm{d}})|0\rangle&=\left\{\iint\frac{1}{4\nu^2}\,\delta(|\boldsymbol{f}|-\nu)NA_{\mathrm{col}}(\boldsymbol{f},\nu)e^{i2\pi\{\boldsymbol{f}\cdot(\boldsymbol{x}_{\mathrm{d}}-\boldsymbol{x})-\nu(t_{\mathrm{d}}-t)\}}d^3\boldsymbol{f}d\nu\right\}^*\\&=\{U_{\mathrm{vac}}(\boldsymbol{x}_{\mathrm{d}}-\boldsymbol{x},t_{\mathrm{d}}-t)\}^*.\end{aligned}\quad(47)$$

This equation is used in category C3. We define the pupil function for the vacuum field $\mathcal{P}_{\text{col}}(\boldsymbol{f},\nu)$ as

$$\mathcal{P}_{\text{col}}(\boldsymbol{f},\nu)=\frac{1}{4\nu^2}\,\delta(|\boldsymbol{f}|-\nu)NA_{\text{col}}(\boldsymbol{f},\nu), \tag{48}$$

$$U_{\text{vac}}(\boldsymbol{x},t)=\int\mathcal{P}_{\text{col}}(\boldsymbol{f},\nu)\,e^{i2\pi(\boldsymbol{f}\cdot\boldsymbol{x}-\nu t)}\,d^3\boldsymbol{f}d\nu. \tag{49}$$

Figure 6 illustrates both $V(\boldsymbol{f},\nu)$ and $\mathcal{P}_{\text{col}}(\boldsymbol{f},\nu)$.

## Observation function

We define the detector function $D(\boldsymbol{x}_{\text{d}},t_{\text{d}})=|D_{\text{R}}(\boldsymbol{x}_{\text{d}},t_{\text{d}})|^2$ to express a detecting plane with uniform photosensitivity, where $D_{\text{R}}(\boldsymbol{x}_{\text{d}},t_{\text{d}})$ is defined as the observation function, which has a random phase distribution (*7*). The detector function is illustrated in Fig. 7. Note that $D_{\text{R}}(\boldsymbol{x}_{\text{d}},t_{\text{d}})$ is a complex function with phase randomness at each space-time point $(\boldsymbol{x}_{\text{d}},t_{\text{d}})$ and a uniformity of modulus, which implies that light fields arriving at the different space-time points do not interfere. The observation function can be written as the observation position function $X_{\text{R}}(\boldsymbol{x}_{\text{d}})$ multiplied by the observation time function $T_{\text{R}}(t_{\text{d}})$ without loss of generality, i.e., $D_{\text{R}}(\boldsymbol{x}_{\text{d}},t_{\text{d}})=X_{\text{R}}(\boldsymbol{x}_{\text{d}})T_{\text{R}}(t_{\text{d}})$, where $X_{\text{R}}(\boldsymbol{x}_{\text{d}})$ and $T_{\text{R}}(t_{\text{d}})$ also have phase randomness. Because of this phase randomness, the observation function has the following fundamental formula:

$$|D_{\text{R}}(\boldsymbol{x}_{\text{d}},t_{\text{d}})|^2 f(\boldsymbol{x}_{\text{d}},t_{\text{d}})g^*(\boldsymbol{x}_{\text{d}},t_{\text{d}})=\int f(\boldsymbol{x}_{\text{d}},t_{\text{d}})D_{\text{R}}^*(\boldsymbol{x}_{\text{d}},t_{\text{d}})g^*(\boldsymbol{x}_{\text{l}},t_{\text{l}})D_{\text{R}}(\boldsymbol{x}_{\text{l}},t_{\text{l}})d\boldsymbol{x}_{\text{l}}dt_{\text{l}}\,. \tag{50}$$

We use the formula in this equation in category C3.

## Operator ordering rule

An operator ordering rule is defined to ensure that the observable image is Hermitian. We perform three steps: before, at the time of, and after the expansion of the square of the modulus in the image formation formula.

① Before the expansion, the time-ordered product is applied.

② At the time of the expansion, the Hermitian conjugate is placed on the right side as shown: $|\hat{O}|^2=\hat{O}\hat{O}^\dagger$.

(Exception: if the vacuum exists on the left side of the diagram, then $|\hat{O}|^2=\hat{O}^\dagger\hat{O}$.)

③ After the expansion, the normal ordered product is applied only to the real-field operators.

The operators used to express the real field and the vacuum field are exchangeable. All operators (q-numbers) can be converted into c-numbers by acting on the vacuum state $|0\rangle$.

## Image formation formula

Here, we derive the image formation formula by using all the knowledge described above. During the derivation process, all fields are expressed as operators, which can be rearranged based on the ordering rule, and then the images for categories C1–C4 can be calculated by acting on $\langle 0|\cdots|0\rangle$ on both sides. The initial image is given by the four-dimensional space-time function, where both $t_{\text{d}}$ and $\tau$ are time-related variables:

$$I(-\boldsymbol{x}',t_{\text{d}},\tau)$$
$$=\langle 0|\int\left|-e^{i\theta}\,\hat{a}_{\text{lo}}(\boldsymbol{x}_{\text{d}},t_{\text{d}}-\tau)+\int\iiint N(\boldsymbol{x}'+\boldsymbol{x})\hat{p}^{(i)}(\boldsymbol{x},t)\hat{a}_{\text{sig}}^\dagger(\boldsymbol{x},\,t)\hat{a}_{\text{col}}(\boldsymbol{x}_{\text{d}},t_{\text{d}})d^3\boldsymbol{x}dt\right|^2 D(\boldsymbol{x}_{\text{d}},t_{\text{d}})d\boldsymbol{x}_{\text{d}}|0\rangle$$

$$= \langle 0| \int \Big| -e^{i\theta}\, \hat{a}_{\mathrm{lo}}(\boldsymbol{x}_{\mathrm{d}}, t_{\mathrm{d}} - \tau) D_{\mathrm{R}}(\boldsymbol{x}_{\mathrm{d}}, t_{\mathrm{d}}) + \int \iiint N(\boldsymbol{x}' + \boldsymbol{x}) \hat{p}^{(i)}(\boldsymbol{x}, t) \hat{a}^{\dagger}_{\mathrm{sig}}(\boldsymbol{x},\, t) \hat{a}_{\mathrm{col}}(\boldsymbol{x}_{\mathrm{d}}, t_{\mathrm{d}}) D_{\mathrm{R}}(\boldsymbol{x}_{\mathrm{d}}, t_{\mathrm{d}}) d^3\boldsymbol{x} dt \Big|^2 d\boldsymbol{x}_{\mathrm{d}} |0\rangle. \quad (51)$$

Note here that $D(\boldsymbol{x}_{\mathrm{d}}, t_{\mathrm{d}})$ is not a function of $t_{\mathrm{d}}$, but $D_{\mathrm{R}}(\boldsymbol{x}_{\mathrm{d}}, t_{\mathrm{d}})$ is a function of $t_{\mathrm{d}}$. Using this equation, we can perform the calculations required for each category.

### 1. *Category C1*

The imaging systems in this category are microscopy systems that use coherent interactions and have an LO, including OCT systems. The image is given by:

$$I_{\mathrm{C1}}(-\boldsymbol{x}', t_{\mathrm{d}}, \tau)$$

$$= \langle 0| \int |\hat{a}_{\mathrm{lo}}(\boldsymbol{x}_{\mathrm{d}}, t_{\mathrm{d}} - \tau)|^2 D(\boldsymbol{x}_{\mathrm{d}}, t_{\mathrm{d}}) d\boldsymbol{x}_{\mathrm{d}}\, |0\rangle$$

$$+ e^{i\theta} \langle 0| \iint \iiint N(\boldsymbol{x}' + \boldsymbol{x}) \hat{p}^{(i)}(\boldsymbol{x}, t) \hat{a}^{\dagger}_{\mathrm{sig}}(\boldsymbol{x},\, t) \hat{a}_{\mathrm{col}}(\boldsymbol{x}_{\mathrm{d}}, t_{\mathrm{d}}) \hat{a}^{\dagger}_{\mathrm{lo}}(\boldsymbol{x}_{\mathrm{d}}, t_{\mathrm{d}} - \tau) D(\boldsymbol{x}_{\mathrm{d}}, t_{\mathrm{d}}) d^3\boldsymbol{x} dt d\boldsymbol{x}_{\mathrm{d}}\, |0\rangle + c.c.$$

$$\approx \int |L(\boldsymbol{x}_{\mathrm{d}}, t_{\mathrm{d}} - \tau)|^2\, D(\boldsymbol{x}_{\mathrm{d}}) d\boldsymbol{x}_{\mathrm{d}}$$

$$+ e^{i\theta}\, T^*_{\mathrm{lo}}(t_{\mathrm{d}} - \tau) \iiint N(\boldsymbol{x}' + \boldsymbol{x}) \int p^{(i)}(\boldsymbol{x}, t) \int ASF_{\mathrm{col}}(\boldsymbol{x}_{\mathrm{d}} - \boldsymbol{x}, t_{\mathrm{d}} - t) X^*_{\mathrm{lo}}(\boldsymbol{x}_{\mathrm{d}}) D(\boldsymbol{x}_{\mathrm{d}}) d\boldsymbol{x}_{\mathrm{d}}\, dt\, d^3\boldsymbol{x} + c.c.$$

$$= Lo(\, t_{\mathrm{d}} - \tau) + e^{i\theta} T^*_{\mathrm{lo}}(t_{\mathrm{d}} - \tau) \iiint N(\boldsymbol{x}' + \boldsymbol{x}) h_{\mathrm{t}}(\boldsymbol{x}, t_{\mathrm{d}}) d^3\boldsymbol{x} + c.c. \quad (52)$$

where the amplitude for the LO is written as $L(\boldsymbol{x}_{\mathrm{d}}, t_{\mathrm{d}} - \tau) = X_{\mathrm{lo}}(\boldsymbol{x}_{\mathrm{d}}) T_{\mathrm{lo}}(t_{\mathrm{d}} - \tau)$ when the detector is placed at $z_{\mathrm{d}} = 0$, and $p^{(i)}(\boldsymbol{x}, t) = \langle 0|\hat{p}^{(i)}(\boldsymbol{x}, t)|0\rangle$. $T_{\mathrm{lo}}(t_{\mathrm{d}})$ corresponds to the Fourier transform of $S_{\mathrm{lo}}(\nu_{\mathrm{d}})$. Here, we omit the electric field dimension of the LO, $E_{\mathrm{lo}}$, from the expression $E_{\mathrm{lo}} X_{\mathrm{lo}}(\boldsymbol{x}_{\mathrm{d}}) T_{\mathrm{lo}}(t_{\mathrm{d}} - \tau)$, but we assume that $E_{\mathrm{lo}}$ is included in $T_{\mathrm{lo}}(t_{\mathrm{d}} - \tau)$.

In the Kohler illumination system, we assume that $D(\boldsymbol{x}_{\mathrm{d}}) = \delta(\boldsymbol{x}_{\mathrm{d}})$, and in this case, the LO corresponds to

$$\hat{a}_{\mathrm{lo}}(\boldsymbol{0}, t_{\mathrm{d}}) \to \int \hat{a}_{\mathrm{ill}}(\boldsymbol{x}, t)\, \hat{a}^{\dagger}_{\mathrm{sig}}(\boldsymbol{x},\, t) \hat{a}_{\mathrm{col}}(\boldsymbol{0}, t_{\mathrm{d}}) d^3\boldsymbol{x} dt, \quad (53)$$

which has the following relation:

$$\langle 0|\hat{a}_{\mathrm{lo}}(\boldsymbol{0}, t_{\mathrm{d}})|0\rangle = \int K(\boldsymbol{x}, t) ASF_{\mathrm{col}}(-\boldsymbol{x}, t_{\mathrm{d}} - t)\, d^3\boldsymbol{x} dt = T_{\mathrm{lo}}(\, t_{\mathrm{d}} - \tau) \int K(\boldsymbol{x}, t) ASF_{\mathrm{col}}(-\boldsymbol{x}, 0)\, d^3\boldsymbol{x} dt. \quad (54)$$

The image in the Kohler illumination system then becomes

$$I_{\mathrm{K}}(-\boldsymbol{x}', t_{\mathrm{d}}, \tau) = Const. + e^{i\theta}\, T^*_{\mathrm{lo}}(t_{\mathrm{d}} - \tau) \int \iiint N(\boldsymbol{x}' + \boldsymbol{x}) K_{\mathrm{ill}}(\boldsymbol{x}, t) ASF_{\mathrm{col}}(-\boldsymbol{x}, t_{\mathrm{d}} - t) d^3\boldsymbol{x} dt + c.c., \quad (55)$$

where

$$K_{\mathrm{ill}}(\boldsymbol{x}, t) = \int S_{\mathrm{ill}}(\nu) |\mathbb{P}_{\mathrm{ill}}(\boldsymbol{f}, \nu)|^2 P^*_{\mathrm{col}}(\boldsymbol{f}, \nu)\, e^{i2\pi(\boldsymbol{f}\cdot\boldsymbol{x} - \nu t)} d^3\boldsymbol{f} d\nu. \quad (56)$$

In the Kohler illumination system, $T_{\mathrm{lo}}(\, t_{\mathrm{d}})$ corresponds to the Fourier transform of $S_{\mathrm{ill}}(\nu)$.

### 2. *Category C2*

The imaging systems in this category are microscopy systems that use coherent interactions and do not have an LO. The image can be written as:

$$I_{\mathrm{C2}}(-\boldsymbol{x}', t_{\mathrm{d}}, 0)$$

$$= \langle 0| \int \left| \iint \iiint N(\boldsymbol{x}' + \boldsymbol{x}) \hat{p}^{(i)}(\boldsymbol{x}, t) \hat{a}_{\text{sig}}^{\dagger}(\boldsymbol{x}, t) \hat{a}_{\text{col}}(\boldsymbol{x}_{\text{d}}, t_{\text{d}}) D_R^*(\boldsymbol{x}_{\text{d}}, t_{\text{d}}) d^3\boldsymbol{x} dt \right|^2 d\boldsymbol{x}_{\text{d}} |0\rangle$$

$$= \langle 0| \left| \iint \iiint N(\boldsymbol{x}' + \boldsymbol{x}) \hat{p}^{(i)}(\boldsymbol{x}, t) \hat{a}_{\text{sig}}^{\dagger}(\boldsymbol{x}, t) \int \hat{a}_{\text{col}}(\boldsymbol{x}_{\text{d}}, t_{\text{d}}) X_R^*(\boldsymbol{x}_{\text{d}}) d\boldsymbol{x}_{\text{d}} \; d^3\boldsymbol{x} dt \right|^2 |0\rangle$$

$$= \left| \iint \iiint N(\boldsymbol{x}' + \boldsymbol{x}) p^{(i)}(\boldsymbol{x}, t) \int ASF_{\text{col}}(\boldsymbol{x}_{\text{d}} - \boldsymbol{x}, t_{\text{d}} - t) X_R^*(\boldsymbol{x}_{\text{d}}) d\boldsymbol{x}_{\text{d}} \; d^3\boldsymbol{x} dt \right|^2$$

$$= \left| \iiint N(\boldsymbol{x}' + \boldsymbol{x}) h_{\text{t}}'(\boldsymbol{x}, t_{\text{d}}) \, d^3\boldsymbol{x} \right|^2 . \tag{57}$$

### *3. Category C3*

The imaging systems in this category are microscopy systems that use incoherent interactions and have an LO. One of the excitation fields considered in this case is the vacuum field, which means that $\hat{p}^{(i)}(\boldsymbol{x}, t)$ includes $\hat{a}_{\text{vac}}(\boldsymbol{x}, t)$. The vacuum field also acts as the LO with $\tau = 0$: $\hat{a}_{\text{lo(v)}}(\boldsymbol{x}_{\text{d}}, t_{\text{d}})$. The image is given by:

$$I_{\text{C3}}(-\boldsymbol{x}', t_{\text{d}}, 0)$$

$$= \langle 0| \int |\hat{a}_{\text{lo}}(\boldsymbol{x}_{\text{d}}, t_{\text{d}}) D_{\text{R}}(\boldsymbol{x}_{\text{d}}, t_{\text{d}})|^2 d\boldsymbol{x}_{\text{d}} \, |0\rangle$$

$$+ i\langle 0| \iint \iiint N(\boldsymbol{x}' + \boldsymbol{x}) \hat{p}^{(i)}(\boldsymbol{x}, t) \hat{a}_{\text{lo}}^{\dagger}(\boldsymbol{x}_{\text{d}}, t_{\text{d}}) D_{\text{R}}(\boldsymbol{x}_{\text{d}}, t_{\text{d}}) \hat{a}_{\text{sig}}^{\dagger}(\boldsymbol{x}, t) \hat{a}_{\text{col}}(\boldsymbol{x}_{\text{d}}, t_{\text{d}}) D_R^*(\boldsymbol{x}_{\text{d}}, t_{\text{d}}) d^3\boldsymbol{x} dt d\boldsymbol{x}_{\text{d}} \, |0\rangle + c.c.$$

$$= \langle 0| \int |\hat{a}_{\text{lo}}(\boldsymbol{x}_{\text{d}}, t_{\text{d}}) D_{\text{R}}(\boldsymbol{x}_{\text{d}}, t_{\text{d}})|^2 d\boldsymbol{x}_{\text{d}} \, |0\rangle$$

$$+ i\langle 0| \int \iiint N(\boldsymbol{x}' + \boldsymbol{x}) \left\{ \int \hat{p}^{(i)}(\boldsymbol{x}, t) \hat{a}_{\text{lo}}^{\dagger}(\boldsymbol{x}_{\text{l}}, t_{\text{l}}) D_{\text{R}}(\boldsymbol{x}_{\text{l}}, t_{\text{l}}) d\boldsymbol{x}_{\text{l}} dt_{\text{l}} \right\}$$

$$\times \left\{ \int \hat{a}_{\text{col}}(\boldsymbol{x}_{\text{d}}, t_{\text{d}}) \hat{a}_{\text{sig}}^{\dagger}(\boldsymbol{x}, t) D_R^*(\boldsymbol{x}_{\text{d}}, t_{\text{d}}) d\boldsymbol{x}_{\text{d}} \right\} d^3\boldsymbol{x} dt |0\rangle$$

$$+ c.c.$$

$$= i \iiint N(\boldsymbol{x}' + \boldsymbol{x}) \int E_{\text{ex}}(\boldsymbol{x}, t) \int \{U_{\text{vac}}(\boldsymbol{x}_{\text{l}} - \boldsymbol{x}, t_{\text{l}} - t)\}^* X_{\text{R}}(\boldsymbol{x}_{\text{l}}) T_{\text{R}}(t_{\text{l}}) d\boldsymbol{x}_{\text{l}} dt_{\text{l}}$$

$$\times \int ASF_{\text{col}}(\boldsymbol{x}_{\text{d}} - \boldsymbol{x}, t_{\text{d}} - t) X_R^*(\boldsymbol{x}_{\text{d}}) T_{\text{R}}^*(t_{\text{d}}) d\boldsymbol{x}_{\text{d}} \, dt \, d^3\boldsymbol{x}$$

$$+ c.c.$$

$$= i T_{\text{R}}^*(t_{\text{d}}) \iiint N(\boldsymbol{x}' + \boldsymbol{x}) \int E_{\text{ex}}(\boldsymbol{x}, t) \underbrace{\left\{ \int \{U_{\text{vac}}(\boldsymbol{x}_{\text{l}} - \boldsymbol{x}, t_{\text{l}} - t)\}^* X_{\text{R}}(\boldsymbol{x}_{\text{l}}) T_{\text{R}}(t_{\text{l}}) d\boldsymbol{x}_{\text{l}} dt_{\text{l}} \right\}}_{\{U'_{\text{vac}}(-\boldsymbol{x}, -t)\}^* \text{ : function of } \boldsymbol{x} \text{ and } t}$$

$$\times \underbrace{\left\{ \int ASF_{\text{col}}(\boldsymbol{x}_{\text{d}} - \boldsymbol{x}, t_{\text{d}} - t) X_R^*(\boldsymbol{x}_{\text{d}}) d\boldsymbol{x}_{\text{d}} \right\}}_{ASF'_{\text{col}}(-\boldsymbol{x}, t_{\text{d}} - t) \text{ : function of } \boldsymbol{x} \text{ and } t_{\text{d}} - t} dt \, d^3\boldsymbol{x}$$

$$+ c.c.$$

$$= i T_{\text{R}}^*(t_{\text{d}}) \iiint N(\boldsymbol{x}' + \boldsymbol{x}) \int E_{\text{ex}}(\boldsymbol{x}, t) \{U'_{\text{vac}}(-\boldsymbol{x}, -t)\}^* ASF'_{\text{col}}(-\boldsymbol{x}, t_{\text{d}} - t) dt \, d^3\boldsymbol{x}$$

$$+ c.c.$$

$$= iT_{\mathrm{R}}^{*}(t_{\mathrm{d}}) \iiint N(\boldsymbol{x}' + \boldsymbol{x}) h_{\mathrm{t}}(\boldsymbol{x}, t_{\mathrm{d}}) d^{3}\boldsymbol{x} + c.c., \tag{58}$$

where we assume that the time-related part of the observation function $T_{\mathrm{R}}(t_{\mathrm{d}})$ includes the electric field dimension of the LO (the vacuum field) $E_{\mathrm{lo}}$, and that $E_{\mathrm{ex}}(\boldsymbol{x}, t)$ corresponds to part of the excitation fields other than the vacuum field. Note that $T_{\mathrm{R}}(t_{\mathrm{d}})$ has the same characteristics in terms of phase randomness as $T_{\mathrm{lo}}(t_{\mathrm{d}})$ when the vacuum field acts as the LO. Because the first term $\langle 0| \int |\hat{a}_{\mathrm{lo}}(\boldsymbol{x}_{\mathrm{d}}, t_{\mathrm{d}}) D_{\mathrm{R}}(\boldsymbol{x}_{\mathrm{d}}, t_{\mathrm{d}})|^{2} d\boldsymbol{x}_{\mathrm{d}} |0\rangle$ corresponding to the vacuum field itself is not observed, it simply disappears (*33*, *40*).

### 4. *Category C4*

The imaging systems in this category are microscopy systems that use incoherent interactions and do not have an LO (*43*). One of the excitation fields considered in this category is the vacuum field, which means that $\hat{p}^{(i)}(\boldsymbol{x}, t)$ includes $\hat{a}_{\mathrm{vac}}(\boldsymbol{x}, t)$. The image can be expressed as:

$$I_{\mathrm{C4}}(-\boldsymbol{x}', t_{\mathrm{d}}, 0)$$
$$= \langle 0| \int \left| \iint \iiint N(\boldsymbol{x}' + \boldsymbol{x}) \hat{p}^{(i)}(\boldsymbol{x}, t) \hat{a}_{\mathrm{sig}}^{\dagger}(\boldsymbol{x}, t) \hat{a}_{\mathrm{col}}(\boldsymbol{x}_{\mathrm{d}}, t_{\mathrm{d}}) D_{R}^{*}(\boldsymbol{x}_{\mathrm{d}}, t_{\mathrm{d}}) d^{3}\boldsymbol{x} dt \right|^{2} d\boldsymbol{x}_{\mathrm{d}} |0\rangle$$
$$= \left| \iint \iiint N(\boldsymbol{x}' + \boldsymbol{x}) p^{(i)}(\boldsymbol{x}, t) \int ASF_{\mathrm{col}}(\boldsymbol{x}_{\mathrm{d}} - \boldsymbol{x}, t_{\mathrm{d}} - t) X_{R}^{*}(\boldsymbol{x}_{\mathrm{d}}) d\boldsymbol{x}_{\mathrm{d}} \; d^{3}\boldsymbol{x} dt \right|^{2}$$
$$= \left| \iiint N(\boldsymbol{x}' + \boldsymbol{x}) h_{\mathrm{t}}'(\boldsymbol{x}, t_{\mathrm{d}}) \, d^{3}\boldsymbol{x} \right|^{2}. \tag{59}$$

### 5. *General expression of the image formation formula*

The image formation formulas for categories C1–C4 can be integrated into Eq. (1). Note that in category C1, where the LO comes from the light source, $T(t_{\mathrm{d}})$ corresponds to $T_{\mathrm{lo}}(t_{\mathrm{d}})$, whereas in category C3, where the vacuum field acts as the LO, $T(t_{\mathrm{d}})$ corresponds to the observation time function $T_{\mathrm{R}}(t_{\mathrm{d}})$.

## Four-dimensional aperture calculation rule

By Fourier transforming the instrumental function $h_{\mathrm{t}}(\boldsymbol{x}, t_{\mathrm{d}})$, we then obtain the calculation rule for the 4D aperture $A_{4}(\boldsymbol{f}, \nu/c)$ for each category. Table 1 shows the correspondence rules between the arrows in the diagram and the 4D pupil functions; these rules cover all imaging types with arbitrary detector sizes, where the detector function $D(\boldsymbol{x}_{\mathrm{d}}, t_{\mathrm{d}}) = |D_{\mathrm{R}}(\boldsymbol{x}_{\mathrm{d}}, t_{\mathrm{d}})|^{2}$ and an observation function $D_{\mathrm{R}}(\boldsymbol{x}_{\mathrm{d}}, t_{\mathrm{d}}) = X_{\mathrm{R}}(\boldsymbol{x}_{\mathrm{d}}) T_{\mathrm{R}}(t_{\mathrm{d}})$ are applied. Substitution of the delta function $\delta(\boldsymbol{x}_{\mathrm{d}})$ into the detector function $D(\boldsymbol{x}_{\mathrm{d}}, t_{\mathrm{d}})$ reduces it to a 4D aperture calculation rule for confocal microscopy, as shown in Table 2.

In category C1, the 4D aperture can be defined in imaging systems with a sufficiently small detector size, e.g., confocal microscopy and Kohler illumination microscopy, with a light source with arbitrary spectral width, and in systems with a finite detector size, e.g., nonconfocal microscopy, with a narrowband light source. However, most imaging systems in category C1 meet these conditions. For example, in nonconfocal SRG microscopy, a picosecond laser is usually applied and can be regarded as being a sufficiently narrowband source. In this case, the 4D aperture for nonconfocal microscopy becomes equivalent to that used for confocal microscopy, as shown in Fig. 4C. In contrast, in categories C2–C4, the 4D aperture is always well defined.

**Acknowledgement**
This work was supported by Nikon Corporation (Fukutake), Core Research for Evolutional Science and Technology (JPMJCR2105, Yasuno), and Japan Society for the Promotion of Science (21H01836 (Yasuno), 22K04962 (Makita))

**Author contributions**
N.F. conceived the study, developed the theoretical framework, performed all derivations and calculations, interpreted the results, and wrote the manuscript. N.F. had discussions with Y.Y. and S.M. about OCT, and Y.Y. prompted N.F. to consider the 4D theory. Y.Y. and S.M. modified the manuscript to improve readability. All authors reviewed and edited the manuscript.

**Competing interests**
The authors declare no competing interests. The authors (N. Fukutake, Y. Yasuno, S. Makita) published Japanese patent (Publication #: WO2023/182011) and applied for US patent (Application #: US 18/892126) related to a part of OCT.

**Data and materials availability**
All data needed to evaluate the conclusions in the paper are present in the paper and/or the Supplementary Materials.

**List of Supplementary Materials**
Supplementary Notes 1 to 8
Supplementary Figs. 1 to 3
List of symbols and abbreviations
References

# Figures

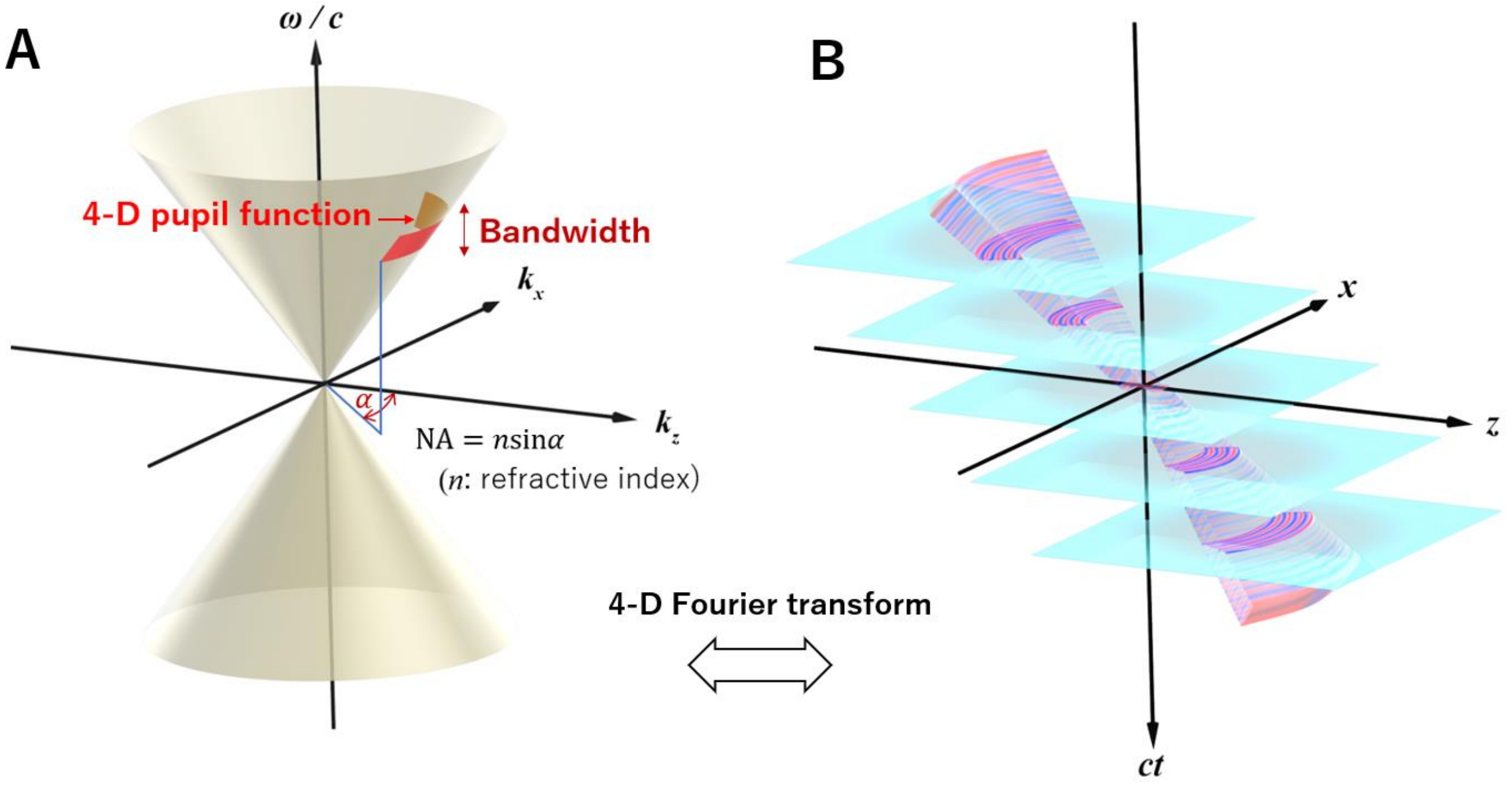

**Fig. 1. Four-dimensional pupil function and four-dimensional amplitude spread function (ASF).** (**A**) Four-dimensional pupil function of an optical system (red surface) defined in a 4D wavevector domain as a part of a conic surface that satisfies the dispersion relation of $k$ and $\omega$. The pupil function width along the $\omega/c$ axis corresponds to the bandwidth of the light, while half of the azimuthal angle-width ($\alpha$) and the NA are related to each other by the equation $NA = n \sin \alpha$. (**B**) Four-dimensional ASF defined in the space-time domain as the 4D Fourier transform of the pupil function. The figure illustrates the real part of the ASF. The blue planes represent time points and are shown to highlight the ASF at these time points. In both figures, $k_y$ and $y$ are omitted for simplicity of visualization.

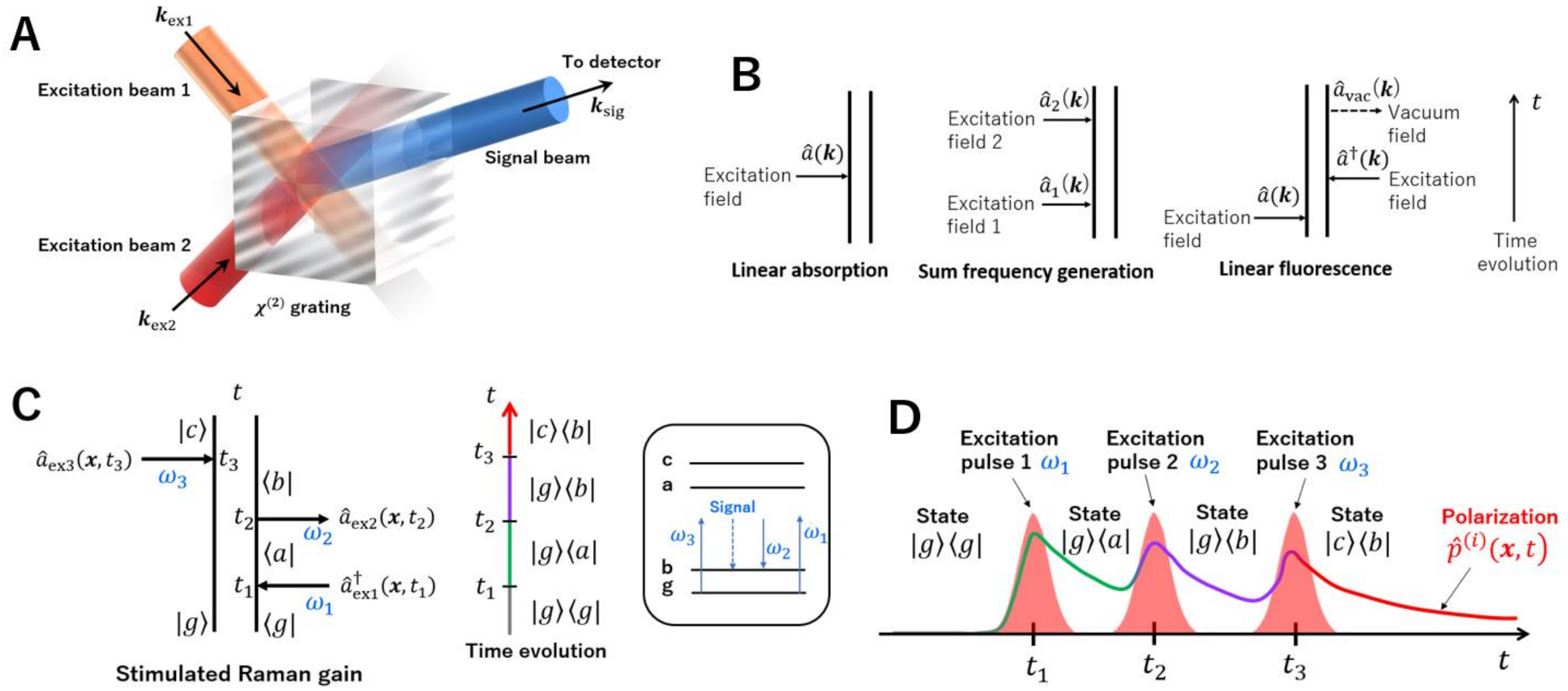

**Fig. 2. Mechanism of the signal generation.** (**A**) Illustration of phase matching (PM). As an example, the PM for sum frequency generation (SFG) involves two excitation beams (orange and red). The SFG signal (blue) is only generated if the wavevectors of these beams ($k_{\mathrm{ex1}}$, $k_{\mathrm{ex2}}$, and $k_{\mathrm{sig}}$) and the $\chi^{(2)}$ grating of the object all satisfy the PM condition. (**B**) Examples of light-matter interactions represented by Feynman diagrams. The arrows represent each excitation field, including the vacuum field. These excitation fields are quantized and repressed as an operator $\hat{a}$. Here, time evolves from the bottom toward the top. (**C**) The 4D operators ($\hat{a}$) representing the excitation fields as functions of space-time. This is exemplified for the SRG case. The bras and kets of a, b, c and g represent the molecule's state and it evolves with each excitation. (**D**) Time evolution of the single-molecule polarization $\hat{p}^{(i)}$ corresponding to the SRG example with $i$ = 3, in C. Red pulses represent excitation lights, and the polarization is pumped by each excitation before decaying. The polarization curve colors represent the molecular states corresponding to the colors in the time evolution diagram in C.

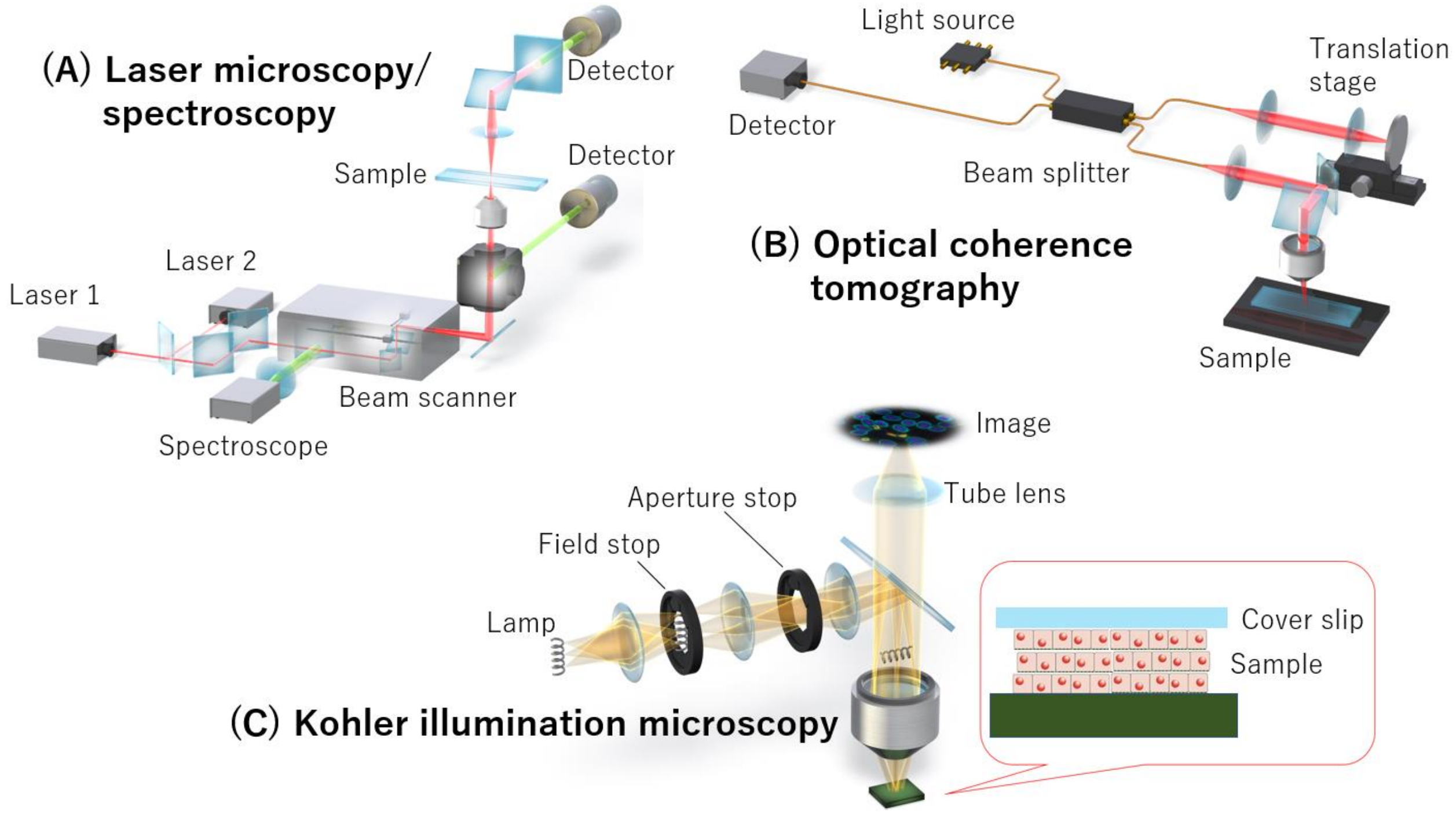

**Fig. 3. Schematics of example imaging systems.** (A) laser scanning microscopy/spectroscopy. (B) optical coherence tomography (OCT). (C) Kohler illumination microscopy. All the optical systems shown are composed of an excitation system, a light-matter interaction area, and a signal correction system. In addition, the OCT system includes a reference arm.

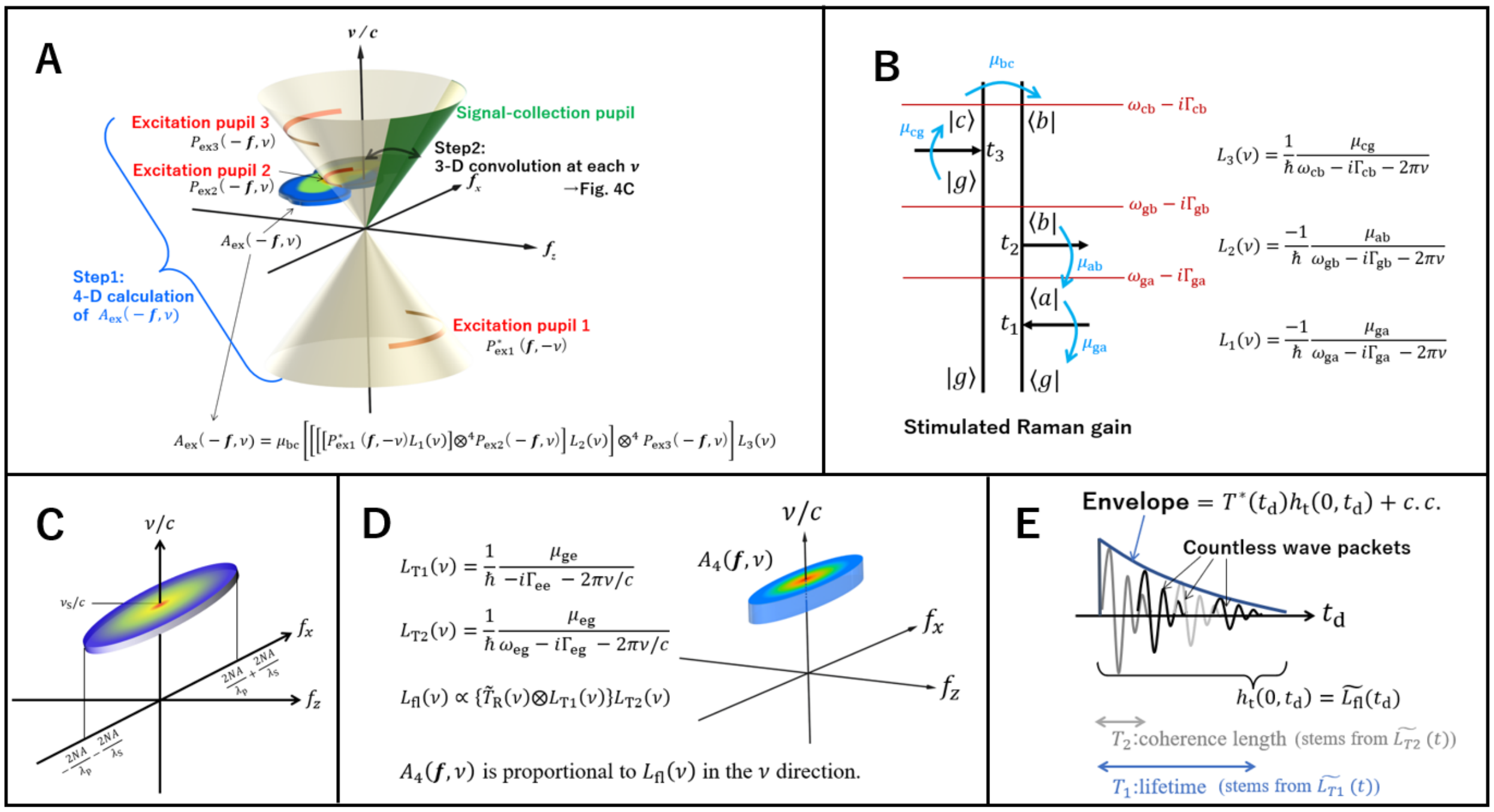

**Fig. 4. Calculation rule for 4D aperture.** (**A**) 4D pupils and 4D aperture in the SRG microscopy case. There are three excitation pupils and one signal collection pupil. The 4D aperture is computed as follows. In step 1, the excitation pupil and Lorentzian functions (see (B)) are combined by 4D convolution and multiplication, as shown in the equation, and $A_{\mathrm{ex}}(-\boldsymbol{f}, \nu)$ is obtained. In step 2, the collection pupil is convolved with $A_{\mathrm{ex}}(-\boldsymbol{f}, \nu)$ in three dimensions at each light frequency to obtain the 4D aperture. (**B**) Calculation rule for the Lorentzian functions. Here, $\omega_{\mathrm{cb}}$ denotes the difference between energy levels c and b, and so on. $\mu_{\mathrm{cg}}$ represents the transition dipole moment from energy level g to level c, and so on. $\Gamma_{ij}$ is the inverse of the transverse relaxation time $T_2$ (for $i \neq j$) or longitudinal relaxation time $T_1$ (for $i = j$). If an arrow is on the left side, the Lorentzian's sign is positive, but if the arrow is on the right side, it becomes negative. (**C**) 4D aperture for confocal SRG microscopy. (**D**) 4D aperture for confocal fluorescence microscopy. (**E**) Temporal profile of the Fourier transform for the 4D aperture, $h_{\mathrm{t}}(\mathbf{0}, t_{\mathrm{d}})$, represented by countless wave packets and a one-sided-exponential-decay envelope.

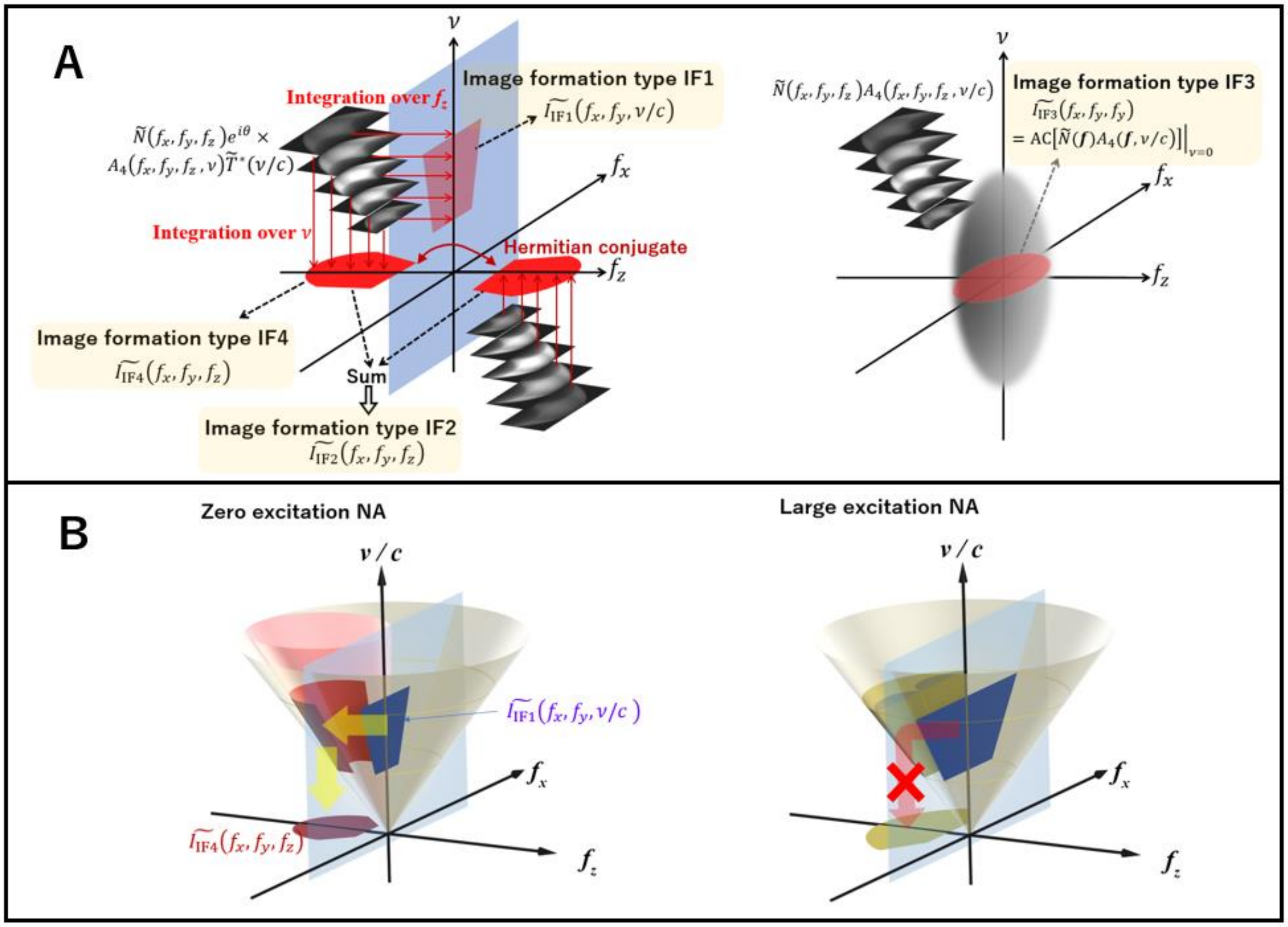

**Fig. 5. Image formation mechanism in 4D frequency space.** (**A**) Four-dimensional frequency-domain representations of image formation types IF1, IF2, IF4 (left), and IF3 (right). $e^{i\theta}B_4(\boldsymbol{f},\nu/c) = e^{i\theta}\tilde{N}(\boldsymbol{f})\{\tilde{T}^*(\nu/c)A_4(\boldsymbol{f},\nu/c)\}$ (gray region) gives the object frequency components windowed by (i.e., multiplied by) the product of the LO frequency component $\tilde{T}^*(\nu/c)$ and the 4D aperture $A_4(\boldsymbol{f},\nu/c)$. The IF1 image is generated by integrating the gray region over $f_z$. For IF2, two gray regions corresponding to $e^{i\theta}B_4(\boldsymbol{f},\nu/c)$ and its Hermitian conjugate are involved. Each region is integrated over $\nu$ to give two frequency components (red plane). The IF2 image is obtained by summing these components and IF4 corresponds to one of them. For simplicity of depiction, these schematics only show specific cases of the reflection type where a single objective lens acts as the excitation and collection lenses. (**B**) Interpretation of refocusing technique using sufficiently thin 4D aperture. The OCT image frequency is in $(f_x, f_y, \nu/c)$-space (blue region). If the 4D aperture has a significant thickness in the $f_z$ direction, then the $f_z$ information is partially lost. Therefore, the OCT image frequency cannot be converted in $(f_x, f_y, f_z)$-space (right). However, if the 4D aperture (red curved surface) is sufficiently thin, the OCT image frequency can be back-projected perfectly.

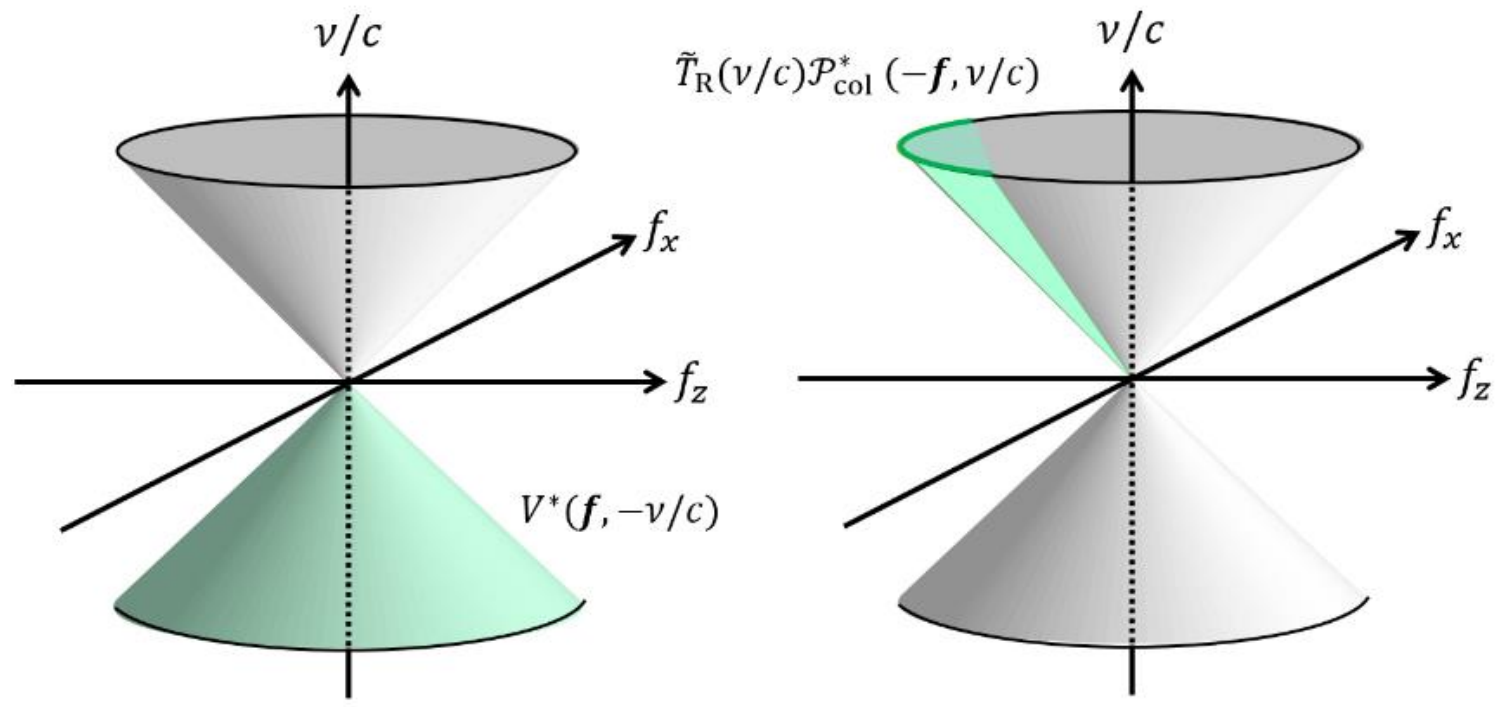

**Fig. 6. Illustration of the pupil function for the vacuum field.**

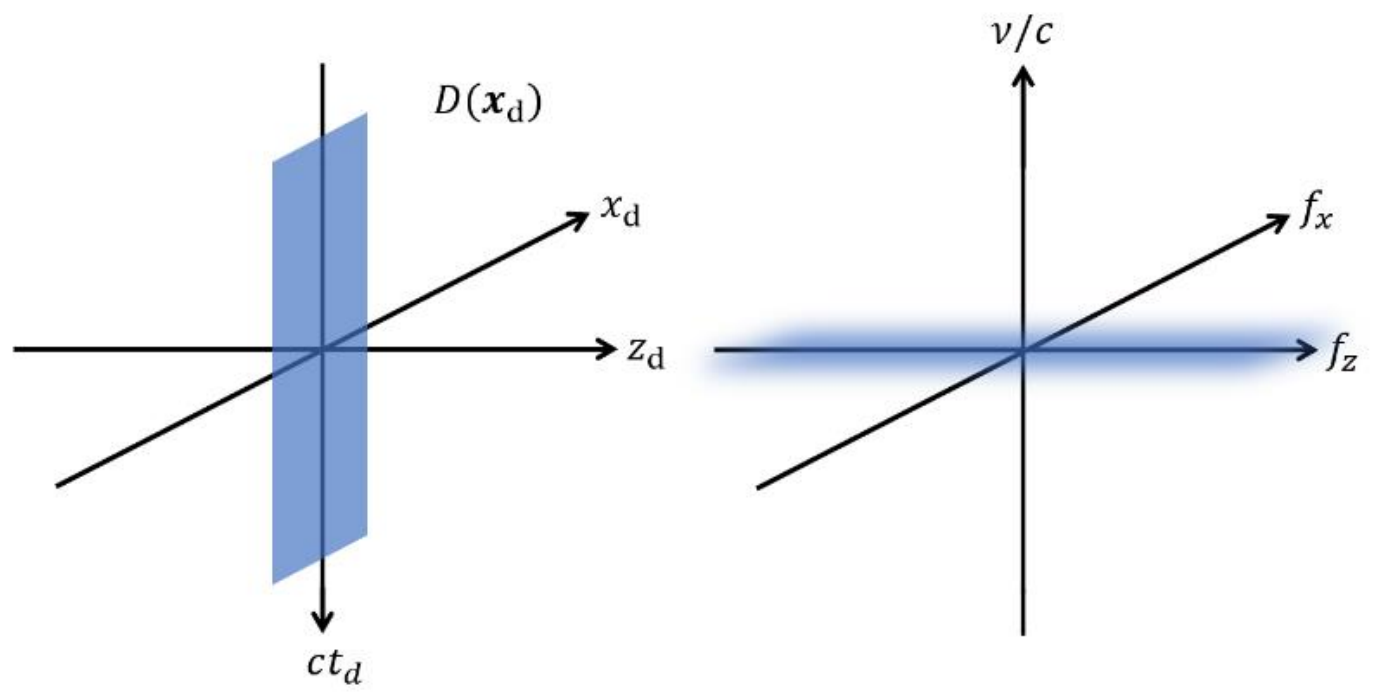

**Fig. 7. Illustration of the detector function in both the 4D frequency domain and the space-time domain.**

**Table 1. Correspondence rules between the arrows in the diagrams and the pupil functions.**

**Category 1)**

| | Excitation | Excitation | Signal |
|---|---|---|---|
| **Arrow** | → | ← | ↜ |
| **Pupil function (laser microscopy)** | $P_{\mathrm{ex}}(-\boldsymbol{f}, \nu/c)$ | $P^*_{\mathrm{ex}}(\boldsymbol{f}, -\nu/c)$ | $P_{\mathrm{col}}(\boldsymbol{f}, \nu/c)\left\{\widetilde{D}(\boldsymbol{f}, \nu/c) \otimes \{\mathbb{P}_{\mathrm{lo}}(\boldsymbol{f}, \nu/c)\}\right\}^*$ |
| **Pupil function (Kohler illumination)** | $S_{\mathrm{ill}}\left(\frac{\nu}{c}\right)\left\|\mathbb{P}_{\mathrm{ill}}\left(-\boldsymbol{f}, \frac{\nu}{c}\right)\right\|^2 P^*_{\mathrm{col}}\left(-\boldsymbol{f}, \frac{\nu}{c}\right)$ | | $P_{\mathrm{col}}(\boldsymbol{f}, \nu/c)$ |

**Category 2)**

| | Excitation | Excitation | Signal |
|---|---|---|---|
| **Arrow** | → | ← | ↜ |
| **Pupil function (laser microscopy)** | $P_{\mathrm{ex}}(-\boldsymbol{f}, \nu/c)$ | $P^*_{\mathrm{ex}}(\boldsymbol{f}, -\nu/c)$ | $P_{\mathrm{col}}(\boldsymbol{f}, \nu/c)\, \tilde{X}^*_R(\boldsymbol{f})$ |
| **Pupil function (Kohler illumination)** | $P_{\mathrm{ill}}(-\boldsymbol{f}, \nu/c)$ | $P^*_{\mathrm{ill}}(\boldsymbol{f}, -\nu/c)$ | $P_{\mathrm{col}}(\boldsymbol{f}, \nu/c)$ |

**Category 3)**

| | Excitation | Excitation | Vacuum | Signal |
|---|---|---|---|---|
| **Arrow** | → | ← | ⇢ | ↜ |
| **Pupil function (laser microscopy)** | $P_{\mathrm{ex}}(-\boldsymbol{f}, \nu/c)$ | $P^*_{\mathrm{ex}}(\boldsymbol{f}, -\nu/c)$ | $\widetilde{D}_{\mathrm{R}}(-\boldsymbol{f}, \nu/c)\mathcal{P}^*_{\mathrm{col}}(-\boldsymbol{f}, \nu/c)$ | $P_{\mathrm{col}}(\boldsymbol{f}, \nu/c)\, \tilde{X}^*_R(\boldsymbol{f})$ |
| **Pupil function (Kohler illumination)** | $P_{\mathrm{ill}}(-\boldsymbol{f}, \nu/c)$ | $P^*_{\mathrm{ill}}(\boldsymbol{f}, -\nu/c)$ | $\tilde{T}_{\mathrm{R}}(\nu/c)\mathcal{P}^*_{\mathrm{col}}(-\boldsymbol{f}, \nu/c)$ | $P_{\mathrm{col}}(\boldsymbol{f}, \nu/c)$ |

**Category 4)**

| | Excitation | Excitation | Vacuum | Signal |
|---|---|---|---|---|
| **Arrow** | → | ← | ⇠ | ↜ |
| **Pupil function** | $P_{\mathrm{ex}}(-\boldsymbol{f}, \nu/c)$ | $P^*_{\mathrm{ex}}(\boldsymbol{f}, -\nu/c)$ | $V^*(\boldsymbol{f}, -\nu/c)$ | $P_{\mathrm{col}}(\boldsymbol{f}, \nu/c)\, \tilde{X}^*_R(\boldsymbol{f})$ |

**Table 2**. **Correspondence rules between arrows and pupil functions for confocal microscopy.**

| | Excitation | Excitation | Vacuum | Vacuum | Signal |
|---|---|---|---|---|---|
| **Arrow** | → | ← | ⇢ | ⇠ | ↜ |
| **Pupil function** | $P_{\mathrm{ex}}(-\boldsymbol{f},\nu/c)$ | $P^*_{\mathrm{ex}}(\boldsymbol{f},-\nu/c)$ | $\tilde{T}_{\mathrm{R}}(\nu/c)\mathcal{P}^*_{\mathrm{col}}(-\boldsymbol{f},\nu/c)$ | $V^*(\boldsymbol{f},-\nu/c)$ | $P_{\mathrm{col}}(\boldsymbol{f},\nu/c)$ |

# Supplementary Information for

## Unified image formation theory for microscopy and optical coherence tomography in 4-D space-time

Naoki Fukutake*, Shuichi Makita, Yoshiaki Yasuno

Corresponding author: Naoki.Fukutake@nikon.com

**The PDF file includes:**

Supplementary Notes 1 to 8
Supplementary Figs. 1 to 3
List of symbols and abbreviations
References

## Supplementary Notes

### Supplementary Note 1: Relationship between 3D aperture and 4D aperture

The 3D aperture $A_3(\boldsymbol{f})$ that was defined in a previous paper (*7*) can be calculated as $A_3(\boldsymbol{f}) = \int \tilde{T}^*(\nu)A_4(\boldsymbol{f},\nu)d\nu$ or $\int A_4(\boldsymbol{f},\nu)d\nu$, depending on the presence or absence of the LO. The 4D aperture confines the signal light frequency and the object's spatial frequencies that contribute to the image formation process, and its Fourier transform shows an aspect of the signal-field amplitude emitted from a single point object into the detector. By Fourier transforming $h_{\mathrm{t}}(\boldsymbol{x},t_{\mathrm{d}})$, we then obtain the 4D aperture calculation rule, which corresponds to a 4D version of the rule from our previous work, i.e., the 3D aperture calculation rule (*7*).

### Supplementary Note 2: Transmission/reflection cross coefficient

We refer to the following relationship between the 3D aperture $A_3(\boldsymbol{f})$ and the transmission cross coefficient (TCC) (*36*, *41*) in a partially coherent imaging system $TCC_3(\boldsymbol{f}_1,\boldsymbol{f}_2)$:

$$TCC_3(\boldsymbol{f}_1,\boldsymbol{f}_2) = A_3(\boldsymbol{f}_1)A_3^*(\boldsymbol{f}_2), \tag{S1}$$

where this relationship holds true for both transmission and reflection type systems. This relationship can be expanded to the 4D TCC by using the 4D aperture $A_4(\boldsymbol{f},\nu)$ as follows:

$$TCC_4(\boldsymbol{f}_1,\nu_1,\boldsymbol{f}_2,\nu_2) = A_4(\boldsymbol{f}_1,\nu_1)A_4^*(\boldsymbol{f}_2,\nu_2). \tag{S2}$$

### Supplementary Note 3: Commutation relation

We show the relationship between the commutation relation (*27*) and the technical terms of optics including ASF and coherence function as follows:

$$\begin{aligned}[\hat{a}_{\mathrm{col}}(\boldsymbol{x}_1,t_1),\hat{a}_{\mathrm{col}}^{+}(\boldsymbol{x}_2,t_2)] &= \iint \frac{1}{2\nu}\ \delta(|\boldsymbol{f}|-\nu)\ |NA_{\mathrm{col}}(\boldsymbol{f},\nu)|^2\ e^{i2\pi[\boldsymbol{f}\cdot(\boldsymbol{x}_1-\boldsymbol{x}_2)-\nu(t_1-t_2)]}d^3\boldsymbol{f}d\nu \\ &= ASF_{\mathrm{col}}^{(\mathrm{na})}(\boldsymbol{x}_1-\boldsymbol{x}_2,t_1-t_2),\end{aligned} \tag{S3}$$

which corresponds to the normalized ASF of the signal collection system with no aberrations, and

$$\begin{aligned}\left[\hat{a}_{\mathrm{ex}}(\boldsymbol{x}_1,t_1),\hat{a}_{\mathrm{ex}}^{\dagger}(\boldsymbol{x}_2,t_2)\right] &= \int \delta(|\boldsymbol{f}|-\nu)\left|\frac{1}{2\nu\alpha_{\boldsymbol{f}}}P_{\mathrm{ex}}(\boldsymbol{f},\nu)\right|^2 e^{i2\pi\{\boldsymbol{f}\cdot(\boldsymbol{x}_1-\boldsymbol{x}_2)-\nu(t_1-t_2)\}}d^3\boldsymbol{f}d\nu \\ &= \gamma_{12}(\boldsymbol{x}_1-\boldsymbol{x}_2,t_1-t_2),\end{aligned} \tag{S4}$$

which corresponds to the 4D coherence function (*40*).

### Supplementary Note 4: Relation between 4D aperture and spectrum/lifetime

As another example of the 4D aperture, we consider a confocal FL microscopy system (*28*, *29*). The 4D theory can be used to provide an appropriate treatment of the vacuum field that influences the spectrum/lifetime measurements in the incoherent interactions. Figure 4D in the main text illustrates the 4D aperture for confocal FL microscopy. The 4D aperture contains information about both the amplitude spectrum of the signal field and the optical resolution. If the spectrum of a single point object $S(\nu) = is(\nu) - is^*(\nu)$ is measured using a spectroscope placed after a confocal pinhole, we obtain $s(\nu) = \tilde{T}^*(\nu) \int A_4(\boldsymbol{f}, \nu)\, d^3\boldsymbol{f}$ by considering each monochromatic wave to interfere with the vacuum field of the same wavelength before detection and using the relation $\int h_{\rm t}(\boldsymbol{0}, t_{\rm d})\, e^{-i2\pi\nu t_{\rm d}} dt_{\rm d} = \int A_4(\boldsymbol{f}, \nu)\, d^3\boldsymbol{f}$. Note that $\tilde{T}(\nu) = \tilde{T}_{\rm R}(\nu)$ in category C3 acts as the amplitude spectrum for the vacuum and is the Fourier transform of $T_{\rm R}(t_{\rm d})$ that corresponds to the time evolution of the vacuum field with random phase when propagating from the vicinity of the object toward the confocal pinhole. The measured spectrum then becomes the real function $S(\nu) = -2{\rm Im}[s(\nu)]$, i.e., a Lorentzian function with a spectral width that is determined by $T_2$. In lifetime measurements with the confocal FL microscopy system, the decay curve of a single point object can be given by the one-sided exponential function $T^*(t_{\rm d}) h_{\rm t}(\boldsymbol{0}, t_{\rm d}) + c.c.$ with lifetime $T_1$, as shown in Fig. 4E in the main text.

## Supplementary Note 5: TD-OCT and FD-OCT

OCT is usually a confocal system that can extract one of the cross terms (see Eq. (52) in Method in the main text). The image from TD-OCT (*37*) can be written as

$$I_{\rm OCT1}(-x', -y', \tau) = -i \int T_{\rm lo}^*(t_{\rm d} - \tau) \iiint N(\boldsymbol{x}' + \boldsymbol{x}) h_{\rm t}(\boldsymbol{x}, t_{\rm d}) d^3\boldsymbol{x}\, dt_{\rm d} \Big|_{z'=0}$$

$$= -i \iiint N(\boldsymbol{x}' + \boldsymbol{x})\, PSF_4(\boldsymbol{x}, \tau) d\boldsymbol{x} \Big|_{z'=0}, \tag{S5}$$

where the observed quantity is integrated over the detection time $t_{\rm d}$. Similarly, the image acquired from frequency-domain OCT (FD-OCT) (*38*) can be expressed as

$$I_{\rm OCT2}(-x', -y', \tau) = -i \int \tilde{T}_{\rm lo}^*(\nu) \iiint N(\boldsymbol{x}' + \boldsymbol{x}) h_\nu(\boldsymbol{x}, \nu) d^3\boldsymbol{x}\, e^{-i\nu\tau} d\nu \Big|_{z'=0}$$

$$= -i \iiint N(\boldsymbol{x}' + \boldsymbol{x})\, PSF_4(\boldsymbol{x}, \tau) d\boldsymbol{x} \Big|_{z'=0}, \tag{S6}$$

where one of the cross terms is extracted from the following expression:

$$\int \left| i\tilde{T}_{\rm lo}(\nu) e^{i\nu\tau'} + \iiint N(\boldsymbol{x}' + \boldsymbol{x}) h_\nu(\boldsymbol{x}, \nu) d^3\boldsymbol{x} \right|^2 e^{-i\nu\Delta\tau} d\nu \Big|_{z'=0}, \tag{S7}$$

where we used the relationship

$$h_\nu(\boldsymbol{x}, \nu) = \int h_{\rm t}(\boldsymbol{x}, t_{\rm d})\; e^{-i\nu t_{\rm d}} dt_{\rm d}. \tag{S8}$$

Note that in FD-OCT, the reference mirror is usually fixed at a position $c\tau'(\neq 0)$ that is shifted from the conjugate plane to the object-coordinate origin $z = 0$ (the geometrical focal plane), i.e., $\tau = \tau' + \Delta\tau$, where $\Delta\tau$ is a variable that emerges from the Fourier transform calculation, which causes an image-origin shift. The images obtained from TD-OCT and FD-OCT, which are denoted by $I_{\rm OCT1}(-x', -y', \tau)$ and $I_{\rm OCT2}(-x', -y', \tau)$, respectively, are identical. Supplementary Fig. 1 shows the 4D aperture and $PSF_4(\boldsymbol{x}, \tau)$ for OCT.

## Supplementary Note 6: Reflection bright-field microscopy

Another example of an image formation type involves a synchronous $(z', \tau)$-scan system that usually operates with an LO from a glass interface in the vicinity of a biological sample (*44*).

In the synchronous $(z', \tau)$-scan system of type IF4, one of the typical microscopy techniques is reflection bright-field microscopy, where the sample stage is scanned in the depth direction along with the glass interface (*44*), as illustrated in Supplementary Fig. 2. For convenience, we divide the illumination pupil function into multiple annular illumination pupil functions at an angle $\phi$ to the optical axis, as shown in Supplementary Fig. 2. Because $\tau$ corresponds to $-2z'\cos\phi$ in the relevant annular illumination, we obtain a 3D image $I_{\mathrm{IF4}}(-x', -y', -z', -2z'\cos\phi)$ that is equivalent to a cross-section of $I_{\mathrm{IF4}}(-\boldsymbol{x}', \tau)$ when cut by the surface $\delta(\tau + 2z'\cos\phi)$. However, because the 3D image formed by this system is a function of $-\boldsymbol{x}'$, the image can be expressed using $I_{\mathrm{IF4}}(-\boldsymbol{x}', -2z') \otimes \delta^3(\boldsymbol{x}')$, which corresponds to the uniform 4D function in the $\tau$ direction; here, $\otimes$ represents a 4D convolution. The Fourier transform of $\{I_{\mathrm{IF4}}(-\boldsymbol{x}', \tau)\delta(\tau + 2z'\cos\phi)\} \otimes \delta(\boldsymbol{x}')$ gives a $\delta(\nu/c)$-plane cross-section of the function $\tilde{I}_{\mathrm{IF4}}(\boldsymbol{f}, \nu/c) = e^{i\theta} B_4(\boldsymbol{f}, \nu/c) + e^{-i\theta} B_4^*(-\boldsymbol{f}, -\nu/c)$ convolved by $\delta(f_x)\delta(f_y)\delta(2\nu\cos\phi/c + f_z)$, and this leads to an image frequency shifted to the origin in the $(f_x, f_y, f_z)$-domain, as described in Supplementary Fig. 2B, where $e^{i\theta}$ is dependent on the refractive index difference between the two sides of the interface. Note that the shift amount varies with the annular radius. By summing all the annular illumination contributions, we then obtain the image frequency of this system, where the image frequency is shifted and distorted when compared with the object frequency. This implies that a restoration of the object frequency occurs with the measurement of the individual annular illuminations.

## Supplementary Note 7: Four-dimensional optical transfer function

For image formation type IF2, in addition to the 4D aperture in the frequency domain, we can define another function for ease of expression of the resolution property: a 4D optical transfer function (OTF). The OTF is used for the light intensity, while the aperture is used for the light amplitude. The OTF is a concept that indicates the ratio of each of the Fourier components in the image and the object. The Fourier components of the image consist of the object's Fourier components alone within the frequency cutoff for the OTF. However, in other imaging systems, the OTF cannot be defined. This implies that new Fourier components that do not exist in the original object may be generated in the image. The OTF can only be defined in $z'$-scan imaging systems with an LO with $\tau = 0$. In type IF2, because of the presence of the LO with $\tau = 0$, the OTF exists, whereas in OCT, the OTF cannot be defined because $\tau \neq 0$, and thus a shift-invariant PSF does not exist.

The 4D OTF is similar to a well-known 3D OTF (*22, 36*). This 3D OTF only represents the spatial imaging property, whereas the 4D OTF also represents the time response caused by the light-matter interaction. The 4D OTF is defined as:

$$OTF_4\left(\boldsymbol{f}, \frac{\nu}{c}\right) = e^{i\theta} O_4\left(\boldsymbol{f}, \frac{\nu}{c}\right) + e^{-i\theta} O_4^*\left(-\boldsymbol{f}, -\frac{\nu}{c}\right), \tag{S9}$$

where $O_4(\boldsymbol{f}, \nu/c) = \tilde{T}^*(-\nu/c) \otimes^{\nu} A_4(\boldsymbol{f}, \nu/c)$, $\otimes^{\nu}$ denotes a convolution in terms of the signal field frequency $\nu$, and $\tilde{T}(\nu/c)$ is the frequency distribution of the LO amplitude. In type IF2, $e^{i\theta}$ becomes $i$. The first and second terms in Eq. (S9) are the Fourier transforms of the second and third terms in Eq. (1) in the main text, respectively, with the exception of the parts relating to the object $\tilde{N}(\boldsymbol{f})$. Because Eq. (S9) represents the sum of a pair of Hermitian conjugate functions, the 4D OTF inevitably becomes a Hermitian function. Using the OTF, the Fourier transform of the IF2 image can be expressed using the simple form, $\tilde{I}_{\mathrm{IF2}}(f_x, f_y, f_y) = \tilde{N}(\boldsymbol{f}) OTF_4(\boldsymbol{f}, \nu/c)|_{\nu=0}$, where $\nu = 0$ stems from the $t_{\mathrm{d}}$ integration.

By Fourier transforming the 4D OTF, we define the 4D point spread function (PSF), $T^*(t_{\mathrm{d}}) h_{\mathrm{t}}\left(\boldsymbol{x}', t_{\mathrm{d}}\right) + c.c.$, which is a real function that expresses a 4D impulse response. Supplementary Fig. 3 shows an example of a 4D OTF (left) in the confocal FL microscopy case.

Along the $\nu$ axis, this 4D OTF shows a complex Lorentzian profile where the width is determined by the longitudinal relaxation time $T_1$ of the excited state. Note that because the LO is the vacuum field, $\tilde{T}(\nu/c)$ has a random phase here. The temporal ($t_\mathrm{d}$ axis) profile of the 4D PSF is a one-sided exponential function with decay time $T_1$.

This 4D OTF is similar to the previously demonstrated 3D OTF (7), but the new OTF is more universal when considered from the following perspective. The previous 3-D OTF has two types, with one each for the real and imaginary parts of $\chi^{(i)}$. The relevant type should have been selected based on *a priori* knowledge of the light-matter interaction. In contrast, the new 3D OTF deduced from the 4D OTF is singular and unambiguous, i.e., it is determined automatically by the relevant light-matter interaction.

Note that because the microscopy technique for image formation type IF4 measures the second term of Eq. (1) from the main text, $I_2$, in isolation from its Hermitian conjugate $I_3$, we can measure not only the amplitude of $I_2$ but also its phase. Therefore, the real and imaginary parts of the corresponding $\chi^{(i)}$ can be measured independently. The image frequency then becomes $\tilde{I}_{\mathrm{IF4}}(f_x, f_y, f_y) = e^{i\theta} \int B_4(\boldsymbol{f}, \nu/c) d\,\nu$, which can be calculated from $e^{i\theta} \tilde{N}(\boldsymbol{f}) O_4(\boldsymbol{f}, \nu/c)|_{\nu=0}$.

**Supplementary Note 8: Imaging systems of image formation types IF1–IF4 and interaction categories C1–C4**

| | Category C1 | Category C2 | Category C3 | Category C4 |
|---|---|---|---|---|
| Type IF1 | OCT | N/A | N/A | N/A |
| Type IF2 | SRS, ISHG | N/A | FL, SRS | N/A |
| Type IF3 | N/A | SHG, THG | N/A | SPDC |
| Type IF4 | IM | N/A | N/A | N/A |

IF1: $\tau$-scan system with LO

IF2: $z'$-scan system with LO

IF3: $z'$-scan system without LO

IF4: $(z', \tau)$-scan system with LO

C1: coherent interaction with LO

C2: coherent interaction without LO

C3: incoherent interaction with LO

C4: incoherent interaction without LO

SRS: spontaneous Raman scattering

SPDC: spontaneous parametric down conversion

ISHG: interference SHG

IM: interference microscopy including ISHG

## Supplementary Figures

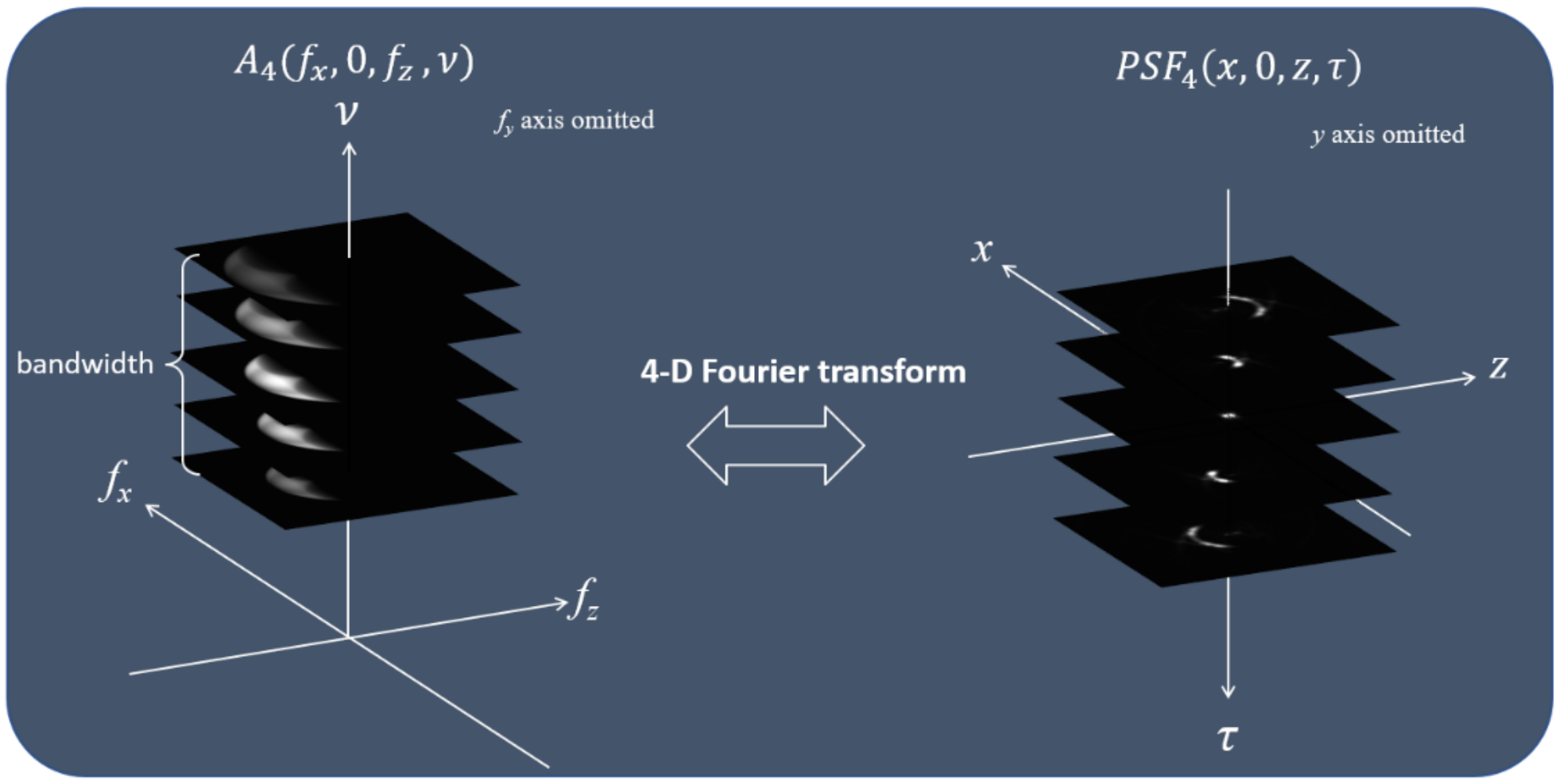

**Supplementary Fig. 1. Calculation example of the 4D aperture for OCT and its Fourier transform.** To exaggerate the characteristics here, the calculation was performed with NA=0.9.

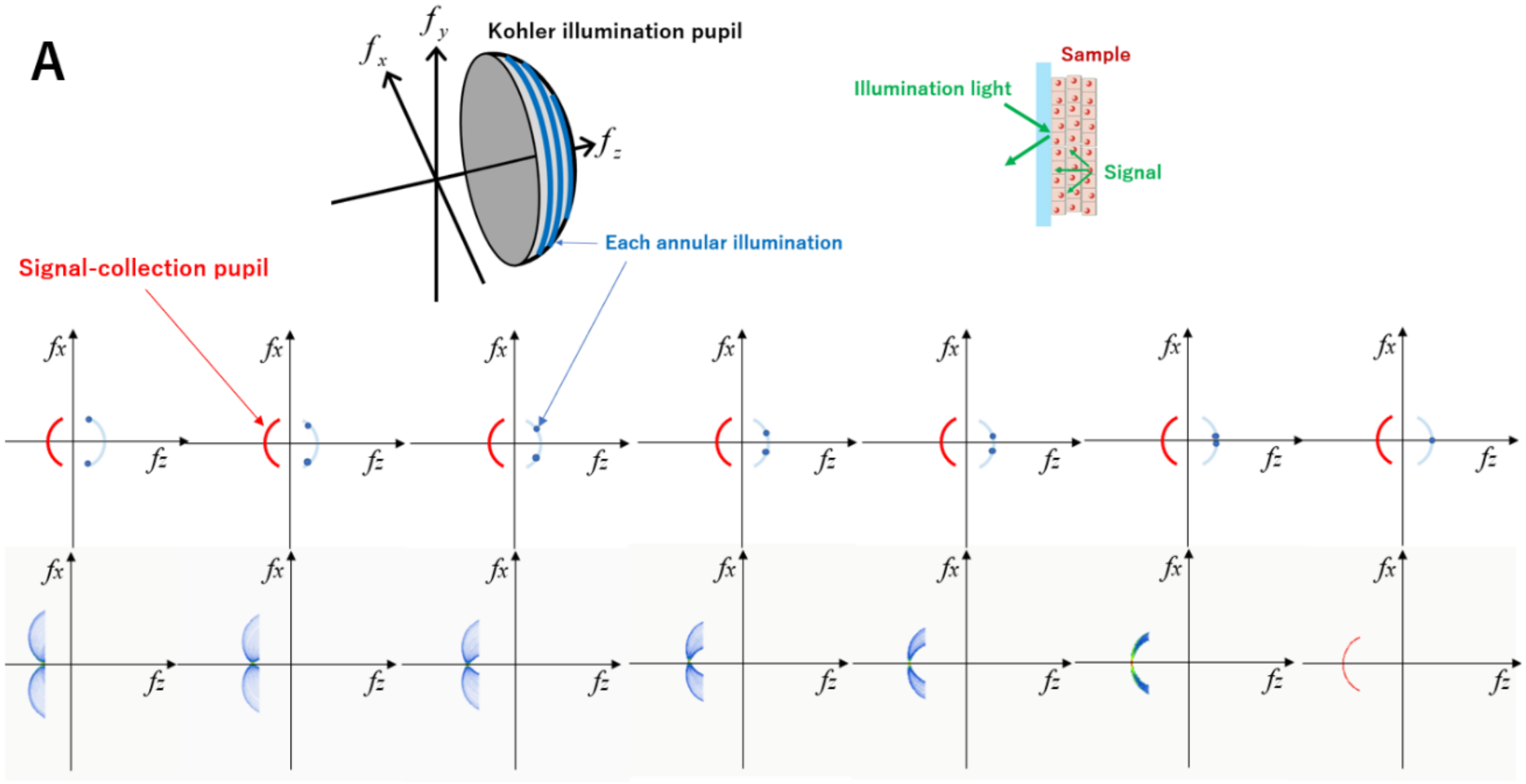

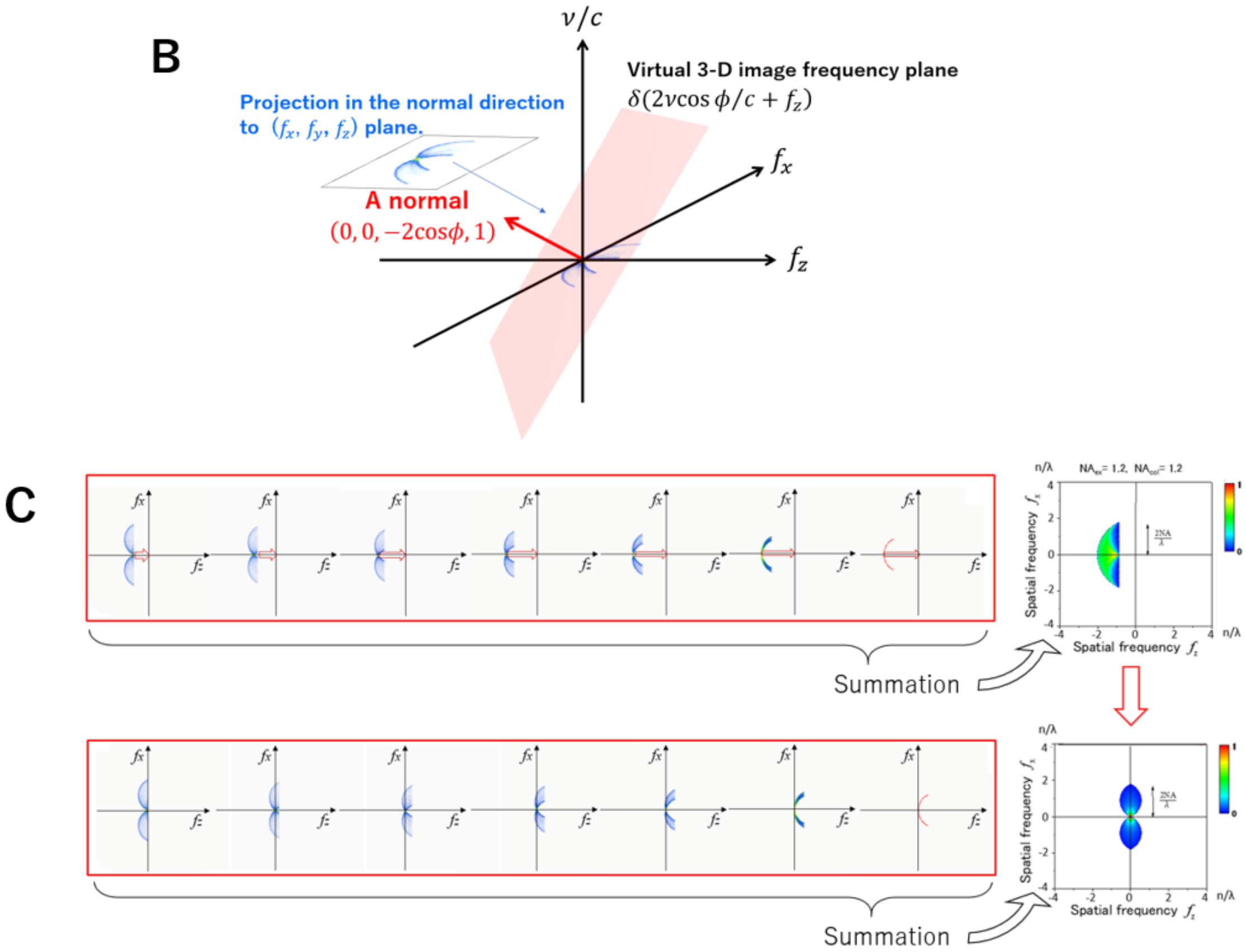

**Supplementary Fig. 2.** (**A**) Image formation process for reflection bright-field microscopy. The Kohler illumination is divided into multiple annular illuminations. (**B**, **C**) Peculiar image formation in reflection bright-field microscopy when the LO is reflected from an interface in the vicinity of the sample. Projection is performed toward the frequency direction corresponding to the real-space direction in which scanning is not performed from the four available directions.

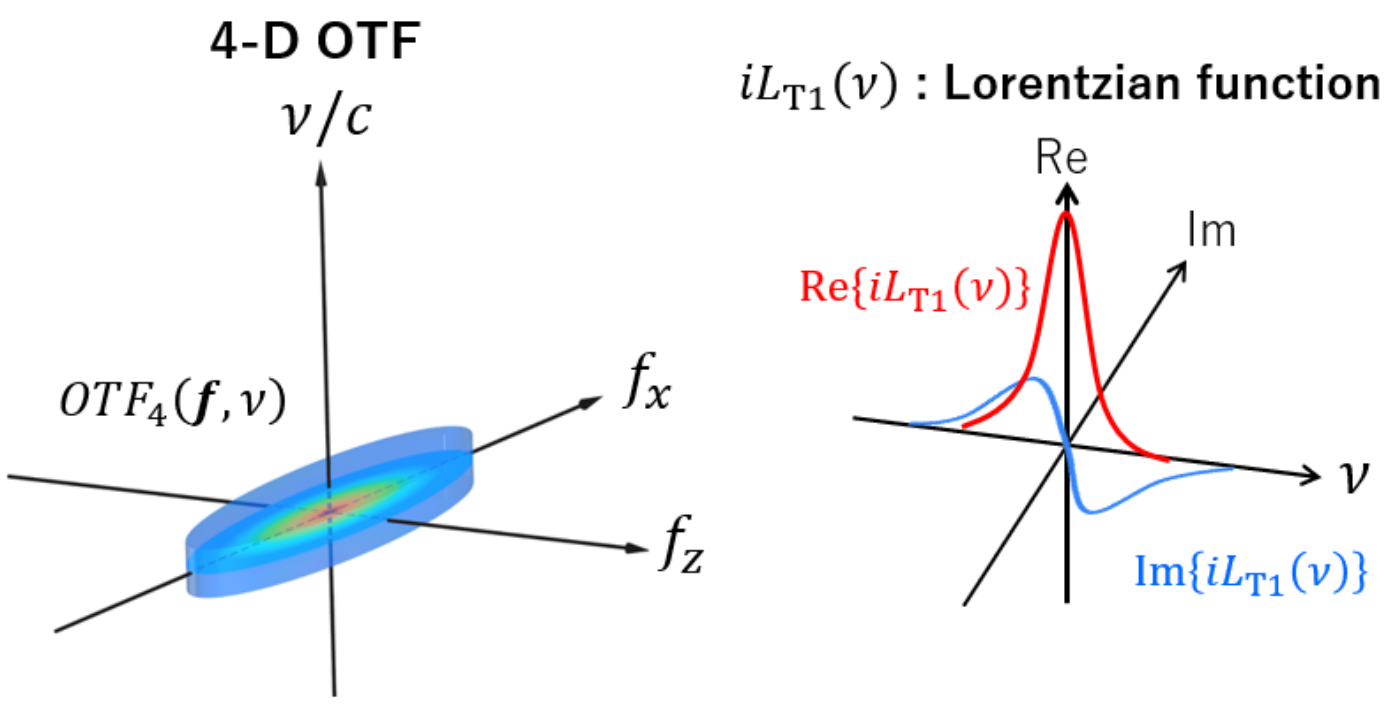

$OTF_4(\boldsymbol{f}, \nu)$ is proportional to $iL_{T1}(\nu)$ in the $\nu$ direction.

**Supplementary Fig. 3. Four-dimensional OTF.** Because of the absence of a carrier structure, the 4D OTF is centered at the zero frequency. Note that in the 4D PSF, which is calculated by Fourier transforming the 4D OTF, the temporal profile consists of only the one-sided exponential decay corresponding to $T_1$.

**List of symbols**

| Symbol | Definition |
|---|---|
| $\lambda$ | Wavelength |
| $\chi^{(i)}$ | *i*-th order susceptibility |
| $\boldsymbol{k} = (k_x, k_y, k_z)$ | 3D wavevector |
| $\omega$ | Angular frequency |
| $c$ | Speed of light |
| $P$ | 4D pupil function |
| $\delta$ | Delta function |
| $h$, *ASF* | 4D amplitude spread function |
| $\boldsymbol{x} = (x, y, z)$ | 3D spatial coordinate |
| $t$ | Time |
| $G_{\mathrm{ret}}$ | Retarded Green's function |
| $\boldsymbol{f} = (f_x, f_y, f_z)$ | 3D spatial frequency |
| $\nu$ | Light frequency |
| $G$ | Grating vector |
| $k_{\mathrm{sig}}$ | Wavevector of the signal wave |
| $k_{\mathrm{ex}}$ | Wavevector of the excitation beam |
| $\hat{P}_{\mathrm{o}}^{(i)}$ | *i*-th order polarization operator |
| $\hat{p}^{(i)}$ | *i*-th order single-molecule polarization |
| $N$ | Molecular density |
| $\chi_{\mathrm{p}}^{(i)}$ | *i*-th order molar susceptibility |
| $\hat{E}_{\mathrm{ex}}^{(i)}$ | Product of all operators representing the excitation fields used to excite the *i*-th interaction |
| $\varepsilon_0$ | Vacuum permittivity |
| $\boldsymbol{x}' = (x', y', z')$ | 3D image spatial coordinate |
| $\tau$ | Delay time of the LO |
| $I$ | Image |
| $t_{\mathrm{d}}$ | Detection time |
| $Lo$ | Intensity of the LO |
| $T$ | Complex amplitude of the LO |
| $\theta$ | Phase shift |
| $h_{\mathrm{t}}$ | Instrumental function |
| $A_4$ | 4D aperture |
| $T_1$ | Longitudinal relaxation time |
| $T_2$ | Transverse relaxation time |
| $D$ | Detector function |
| $D_{\mathrm{R}}$ | Observation function |
| $\boldsymbol{x}_{\mathrm{d}}$ | Detection coordinate |
| $L_i$ | Complex Lorentzian function of *i*-th transition |
| $OTF_4$ | 4D OTF |
| $\otimes^{\nu}$ | Convolution in terms of signal field frequency $\nu$ |
| $\tilde{T}$ | LO frequency distribution |
| $I_i$ | Image of type *i* |
| $\mathrm{AC}$ | Autocorrelation |
| $\phi$ | Illumination angle relative to the optical axis |
| $\delta^3$ | 3D delta function |
| $\hbar$ | Reduced Planck constant |
| $\hat{D}$ | Displacement operator |
| $\lvert 0\rangle$ | Vacuum state |
| $\hat{a}_{\mathrm{EX}}$ | Excitation field in the object |
| $NA_{\mathrm{ex}}$ | NA of the excitation system |
| $p_{\mathrm{ex}}$ | Normalized spectrum |
| $\Theta$ | Heaviside step function |
| $P_{\mathrm{ex}}$ | 4D pupil function in the excitation system |
| $ASF_{\mathrm{ex}}$ | ASF formed by the excitation field in the object |
| $\lvert\alpha_{\boldsymbol{f}}\rangle$ | Coherent state |
| $\lvert\ \ \rangle$ | State |
| $\hat{O}$ | Operator |
| $S$ | Light source spectrum |
| $\hat{a}_{\mathrm{ex}}$ | Annihilation operator for excitation field |
| $\lvert c_{\boldsymbol{f}}\rangle$ | Kohler illumination state |
| $c(\alpha_{\boldsymbol{f}})$ | Coefficient |
| $\hat{a}_{\mathrm{ill}}$ | Annihilation operator |
| $P_{\mathrm{ill}}$ | 4D pupil function of the Kohler illumination system |
| $R$ | Function with a random phase |
| $K$ | Light amplitude distribution in the object |
| $\Gamma_{12}$ | Mutual intensity |
| $t_i'$ | Time of *i*-th excitation |
| $\omega_{\mathrm{bc}}$ | Difference between energy levels b and c in a molecule |
| $\mu_{\mathrm{bc}}$ | Transition dipole moment from energy level c to energy level b |
| $\hat{a}_{\mathrm{lo}}$ | Operator of ASF formed by the LO |
| $p_{\mathrm{lo}}$ | Normalized spectrum of the LO |
| $NA_{\mathrm{lo}}$ | NA of the LO |
| $\alpha'_{\boldsymbol{f}_{\mathrm{d}}}$ | Average amplitude of the LO |
| $\alpha_{\boldsymbol{f}}$ | Complex number |
| $P_{\mathrm{lo}}$ | 4D pupil function of the LO |
| $S_{\mathrm{lo}}$ | Spectrum of the LO |
| $L$ | Amplitude of the LO |
| $\hat{a}_{\mathrm{vac}}$ | Annihilation operator of the vacuum field in the object |
| $V$ | Random phase function with a uniform modulus |
| $\hat{a}_{\mathrm{lo(v)}}$ | Vacuum field acting as the LO at the detector plane |
| $NA_{\mathrm{col}}$ | 4D function that restricts the NA of the signal collection system and includes the aberration and transmittance |
| $\mathcal{P}_{\mathrm{col}}$ | Pupil function for the vacuum field |
| $P_{\mathrm{col}}$ | 4D pupil function of the signal collection system |
| $X_{\mathrm{R}}$ | Observation position function |
| $T_{\mathrm{R}}$ | Observation time function |
| $A_3$ | 3D aperture |
| $TCC_3$ | 3D transmission cross coefficient |
| $E_{\mathrm{ex}}$ | Portion of the excitation fields other than the vacuum field |
| $TCC_4$ | 4D transmission cross coefficient |
| $I_{\mathrm{OCT1}}$ | TD-OCT signal |
| $I_{\mathrm{OCT2}}$ | FD-OCT signal |
| $ASF_{\mathrm{col}}^{(\mathrm{na})}$ | Normalized ASF of the signal collection system with no aberration |
| $\gamma_{12}$ | 4D coherence function |

**List of abbreviations**

| Abbreviation | Definition |
|---|---|
| NA | Numerical aperture |
| FL | Fluorescence |
| TPEF | Two-photon excited fluorescence |
| SHG | Second harmonic generation |
| THG | Third harmonic generation |
| CRS | Coherent Raman scattering |
| OCT | Optical coherence tomography |
| PP | Pump-probe |
| PM | Phase matching |
| LO | Local oscillator |
| ASF | Amplitude spread function |
| SFG | Sum-frequency generation |
| RA | Reference arm |
| SRG | Stimulated Raman gain |
| OTF | Optical transfer function |
| PSF | Point spread function |
| TCC | Transmission cross coefficient |

**Image formation types**

| Type | *z'* scan | *τ* scan | LO |
|---|---|---|---|
| 1 | No | Yes | Yes |
| 2 | Yes | No | Yes |
| 3 | Yes | No | No |
| 4 | Yes | Yes | Yes |

**Imaging system category**

| Presence of LO / Light-molecule interaction type | With LO | Without LO |
|---|---|---|
| Coherent | Category 1 | Category 2 |
| Incoherent | Category 3 | Category 4 |

**Correspondence table of Fourier transform**

| | |
|---|---|
| $h_t$: Instrumental function | $A_4$: Aperture |
| *h, ASF*: Amplitude spread function | *P*: Pupil function |
| $PSF_4$ | $OTF_4$ |